\documentclass[aps,prb,twocolumn,showpacs,10pt,superscriptaddress,floatfix,longbibliography]{revtex4-2}

\usepackage{newtxtext}
\usepackage{newtxmath}
\usepackage{graphicx}
\usepackage{subfigure}
\usepackage{amsmath}
\usepackage{amsfonts}
\usepackage{mathrsfs}
\usepackage{float}
\usepackage{booktabs}
\usepackage{bbm}
\usepackage{xcolor}
\usepackage[colorlinks=true]{hyperref}
\usepackage{ulem}
\usepackage[T1]{fontenc}
\usepackage{svg}
\usepackage{appendix}

\begin{document}
\title{Semiclassical theory for the orbital magnetic moment of superconducting
quasiparticles}
\author{Jian-Hua Zeng}
\affiliation{School of Physics, Sun Yat-sen University, Guangzhou 510275, China}
\author{Zhongbo Yan}

\affiliation{School of Physics, Sun Yat-sen University, Guangzhou 510275, China}
\affiliation{Guangdong Provincial Key Laboratory of Magnetoelectric Physics and
Devices, Sun Yat-sen University, Guangzhou 510275, China}

\author{Zhi Wang}
\email{wangzh356@mail.sysu.edu.cn}

\affiliation{School of Physics, Sun Yat-sen University, Guangzhou 510275, China}
\affiliation{Guangdong Provincial Key Laboratory of Magnetoelectric Physics and
Devices, Sun Yat-sen University, Guangzhou 510275, China}

\author{Qian Niu}
\affiliation{Department of Physics, University of Science and Technology of China, Anhui, China}

\begin{abstract}
We study the orbital magnetic moment of Bogoliubov quasiparticles in superconductors with the semiclassical approach. We derive the orbital magnetic moment of a quasiparticle wavepacket by considering the energy correction of the wavepacket to the linear order of the magnetic field. The semiclassical result is further verified by a linear response calculation with a full quantum mechanical method. From the analytical expression we find that nontrivial structure in the superconducting pairing gap alone is unable to produce quasiparticle orbital magnetic moment, which is in sharp contrast to the behavior of quasiparticle Berry curvatures. We apply the formula to study a tight-binding model with chiral $d$-wave superconducting gap, and show the influence of orbital magnetic moment on the energy spectrum and local density of states. We also calculate the orbital Nernst effect driven by the interplay between the orbital magnetic
moment and the Berry curvature of Bogoliubov quasiparticles.
\end{abstract}
\maketitle

\section{Introduction}
Orbital magnetic moment plays an important role in the magnetic and optical properties of Bloch electrons~\citep{JPCM2010Resta_OM,IJMPB2011Thonhauser_OM,JPCS2019Aryasetiawan_OMMreview,AIPX2024Atencia_omm}. The modern understanding of the orbital magnetic moment for Bloch electrons was formulated by several distinct methods such as the Wannier function approach~\citep{PRL2005Thonhauser_WannierOM,PRB2006Ceresoli_WannierOM,PRB2008Souza_CD}, the linear response approach~\citep{PRL2007Shi_quantumOM}, and the semiclassical approach~\citep{PRL2005Xiao_SemiOM,PRL2006Xiao_SemiOM}. In the semiclassical picture, the orbital magnetic moment of Bloch electrons originates from the self-rotation of the wavepackets~\citep{PRB1996Chang_electronOMM,PRB1999Sundaram_semiclassical,RMP2010Xiao_BerryPhase}. It is a crucial part of the orbital magnetization of the electron system~\citep{PRL2005Xiao_SemiOM,PRL2006Xiao_SemiOM}, which might be even larger than spin-induced magnetization. The orbital magnetic moment is also directly
linked to various magnetic phenomena such as magnetoresistivity~\citep{PRB2022Lahiri_magnetoresistivity,PRB2023Faridi_magnetoresistivity,PRB2025Tu_magnetoresistivity},
Zeeman splitting~\citep{PRB2015Rostami_Zeeman,PRB2018Knothe_Zeeman,PRL2020Lee_Zeeman},
gyrotropic magnetic effect~\citep{PRL2016Zhong_GME,PRB2020Wang_GME},
and magnetic susceptibility~\citep{PRB2015Gao_susceptibility}. Furthermore,
the orbital magnetic moment plays a pivotal role in transport phenomena, such as
the nonlinear thermoelectric Hall effect~\citep{PRB2024Yamaguchi_NTHE,PRB2025Nakazawa_NTHE}, the valley Hall effect~\citep{PRL2007Xiao_Graphene,PRB2021Bhowal_VHE,PRL2024Das_VHE}, and the orbital Hall effect~\citep{PRB2008Tanaka_OHE,PRL2009Kontani_OHE,PRB2018Jo_OHE,PRL2018Go_OHE,PRM2022Salemi_ONE,Nat2023Choi_OHEexp,PRL2023Igor_OHEexp}.

In superconducting systems, Bogoliubov quasiparticles behave similarly to Bloch electrons with a band structure determined by the Bogoliubov-de Gennes (BdG) Hamiltonian~\citep{Zhu2016BdG}. Since the BdG Hamiltonian is constructed by both the electron Hamiltonian and the superconducting gap function, the structure of the quasiparticle band is shaped by the interplay of the electron Bloch functions and the superconducting gap symmetry \citep{RMP2010Hasan_Topological,RPP2017Sato_Topological,JPSP2016Sato_Topological}. This interplay induces rich topological phases which support exotic boundary modes such as the chiral Majorana mode~\citep{RMP2010Hasan_Topological,RPP2017Sato_Topological,JPSP2016Sato_Topological}. Meanwhile, the quasiparticle band structure also exhibits nontrivial geometric properties such as Berry curvatures~\cite{NC2015Cvetkovic_SCBerry,PRB2015Murray_SCBerry,PRB2017Liang_semiSC,PRL2021Wang_semiSC,PRB2023Zhou_SCBerry,PRB2023Zhang_SCBerry,Arxiv2024liao_semiSC,PRB2025Liao_SCBerry,Arxiv2025Hsieh_SCBerry}, which can induce an anomalous Hall response for quasiparticles even in the topological trivial state~\cite{Arxiv2024liao_semiSC}. 

From the perspective of BdG band, the orbital magnetic moment of superconducting quasiparticles obviously deserves closer investigation. It has already drawn theoretical interest~\cite{NJP2009Annet_SComm,PRB2020Robbins_SComm,PRB2021Xiao_Conserved,PRR2021He_SComm,Arxiv2026Zhu_SCOM}. With a Wannier function approach, the orbital angular momentum of superconducting quasiparticles has been derived~\cite{NJP2009Annet_SComm,PRB2020Robbins_SComm}, and the resulting orbital magnetization is decomposed into local and itinerant contributions. The orbital magnetic moment is also discussed from the perspective of conserved charge current in superconductors~\cite{PRB2021Xiao_Conserved}. Linear response approach has also been taken to investigate the orbital magnetization and calculate various effects such as the orbital Edelstein effect~\cite{PRR2021He_SComm,PRL2022Chirolli_SCOE,PRB2024Ando_SCOE}. Unlike the Bloch electrons, different methods for studying the orbital magnetic moment of superconducting quasiparticles often lead to different results. The main difficulty lies in the charge non-conservation of the mean-field BdG Hamiltonian, which brings a discrepancy between the charge distribution and the probability distribution of a quasiparticle wavepacket. This induces a subtle distinction between the orbital magnetic moment and the orbital angular momentum of superconducting quasiparticles. 

Motivated by this theoretical gap, we provide a derivation of the orbital magnetic moment of Bogoliubov quasiparticles with a semiclassical approach.
For this purpose we construct a quasiparticle wavepacket with the eigenstates of the BdG Hamiltonian, and calculate the gradient correction to the energy of the wavepacket with respect to the magnetic field. Following the protocol for treating Bloch electrons~\citep{PRB1996Chang_electronOMM,PRB1999Sundaram_semiclassical,RMP2010Xiao_BerryPhase}, we define the coefficient of the energy correction with respect to a homogeneous magnetic field as the orbital magnetic moment of the quasiparticle, and find that its expression is not a simple generalization of the formula from Bloch bands to BdG bands. We further verify the semiclassical result with a linear response approach. We implement the formula to study a toy model with a chiral $p$-wave pairing, and find that the chiral pairing alone is unable to produce nonzero orbital magnetic moment, which is in sharp contrast to the Berry curvatures~\cite{PRL2021Wang_semiSC,Arxiv2024liao_semiSC}. We show that the orbital magnetic moment can modulate the energy spectrum and the local density of states. Also, it can induce an orbital Nernst effect in superconductors. 
We study a honeycomb lattice with chiral $d$-wave pairing, and calculate the momentum-space distribution of orbital magnetic moment and the induced spectroscopic and transport responses.

The paper is organized as follows. In Sec.~\ref{sec:Theory}, we
derive the expression for the orbital magnetic moment of the superconducting
quasiparticle in the semiclassical framework. In Sec.~\ref{sec:Quantum}, we
verify the semiclassical result with a linear response approach. In Sec.~\ref{sec:AngularMomentum}, we compare the orbital magnetic moment with the orbital angular momentum. In Sec.~\ref{sec:PhysicalProperties}, we calculate the physical responses induced by the orbital
magnetic moment. In Sec.~\ref{sec:ModelCalculation},
we present numerical studies of a tight-binding model with chiral $d$-wave superconducting gap. In Sec.~\ref{sec:Conclusion}, we give a summary. Details of the semiclassical derivation are given in the Appendix.

\section{\protect\label{sec:Theory} Semiclassical theory for orbital magnetic moment in superconductors}

The semiclassical theory for superconducting quasiparticles mathematically resembles the theory for Bloch electrons~\cite{PRL2021Wang_semiSC,Arxiv2024liao_semiSC}, where the dynamics of the quasiparticle wavepacket is dominated by a local approximation of the Hamiltonian in which the position dependence of the external perturbation fields is approximated by the center position of the wavepacket $\mathbf{r}_{c}$. For the BdG Hamiltonian, the local approximation writes as,
\begin{equation}\label{eq:localH}
\hat{H}_c=\left(\begin{array}{cc}
\hat h_0 [\mathbf{r}_c, \mathbf{ r},\mathbf{p}+e\mathbf{A}(\mathbf{r}_c)] & \hat \Delta({\bf r}_c,\mathbf{p})\\
\hat \Delta^{\dagger}({\bf r}_c,\mathbf{p}) & -\hat h^*_0 [\mathbf{r}_c, \mathbf{r},\mathbf{p}+e\mathbf{A}(\mathbf{r}_c)] 
\end{array}\right),
\end{equation}
where $\hat h_0$ is the electron Hamiltonian, ${\mathbf{p}} = -i\hbar \nabla_{\mathbf r}$ is the electron momentum operator and $-e$ is the electron
charge, $\hat \Delta$ is the superconducting pairing Hamiltonian, and $\mathbf{A}$ is the vector potential which was taken as the external perturbation field for the sake of studying orbital magnetic moment~\citep{PRB1996Chang_electronOMM}. The momentum dependence of the pairing Hamiltonian reflects the pairing symmetry of the Cooper pair that involves the relative position of the two electrons of a Cooper pair~\cite{Arxiv2024liao_semiSC}, therefore it does not depend on the vector potential. This local Hamiltonian has the same spatial periodicity as the homogeneous bulk Hamiltonian, therefore it satisfies the eigenvalue equation
\begin{equation}
\hat{H}_{c}\left|\psi_{n,\mathbf{k},\mathbf{r}_{c}}\right\rangle =E_{c}\left|\psi_{n,\mathbf{k},\mathbf{r}_{c}}\right\rangle ,
\end{equation}
 where $E_c$ is the eigenenergy, $\left|\psi_{n,\mathbf{k},\mathbf{r}_{c}}\right\rangle =e^{i\mathbf{k}\cdot\mathbf{r}}\left|\phi_{n,\mathbf{k},\mathbf{r}_{c}}\right\rangle$
is the eigen function with 
$\left|\phi_{n,\mathbf{k},\mathbf{r}_{c}}\right\rangle$ being the cell-periodic part, $n$ labels the BdG band and $\mathbf k$ denotes the momentum.
We can construct a quasiparticle wavepacket~\citep{PRB1999Sundaram_semiclassical,PRB2017Liang_semiSC,Arxiv2024liao_semiSC}
at the $n$-th BdG band with these eigen functions
\begin{equation}
\left|\Psi_{\mathbf{k}_{c},\mathbf{r}_{c}}\right\rangle =\int d\mathbf{k}\alpha_{\mathbf{k}}\left|\psi_{n,\mathbf{k},\mathbf{r}_{c}}\right\rangle ,
\end{equation}
 where $\alpha_{\mathbf{k}}=\left|\alpha_{\mathbf{k}}\right|e^{-i\gamma_{\mathbf{k}}}$
is the constructing function, and ${\bf k}_{c}=\int d\mathbf{k}|\alpha_{\mathbf{k}}|^{2}\mathbf{k}$
is the center of the wavepacket in the momentum space.
The energy of the wavepacket $E$, evaluated up to the first order in the perturbation gradients, is given by:
\begin{align}
E & \approx\langle\Psi|\hat{H}_{c}|\Psi\rangle+\langle\Psi|\Delta\hat{H}|\Psi\rangle=E_{c}+\Delta E_{n,\mathbf{k}_c},
\end{align}
where $\Delta E_{n,\mathbf{k}_c}$ is the wavepacket averaging over the gradient correction to the local Hamiltonian which writes as~\citep{PRB1999Sundaram_semiclassical}
\begin{equation}
\Delta\hat{H}=\frac{1}{2}\left[(\hat{\mathbf{r}}-\mathbf{r}_{c})\cdot\frac{\partial\hat{H}_{c}}{\partial\mathbf{r}_{c}}+\frac{\partial\hat{H}_{c}}{\partial\mathbf{r}_{c}}\cdot(\hat{\mathbf{r}}-\mathbf{r}_{c})\right],
\end{equation}
where $\mathbf{\hat r} = \mathbf {r}\tau_0$ is the quasiparticle position operator with $\tau_0$ the identity matrix acting on the particle-hole space of the BdG Hamiltonian. 
In semiclassical theory, the orbital magnetic moment resides in the first-order energy corrections to the magnetic field. Therefore, we consider a homogeneous magnetic field $\mathbf{B}$ with a circular gauge $\mathbf{A}(\mathbf{r}_{c})=(\mathbf{B}\times\mathbf{r}_{c})/2$. Then the gradient expansion of the local Hamiltonian writes as, 
\begin{equation}
\Delta\hat{H}=\frac{e}{4m}\mathbf{B}\cdot [(\hat{\mathbf{r}}-\mathbf{r}_{c})\times\mathbf{p}\tau_0-\mathbf{p}\tau_0 \times (\hat{\mathbf{r}}-\mathbf{r}_{c})],
\end{equation}
where $m$ is the mass of the electron. Then the first-order energy correction
from a small homogeneous magnetic field is written as
$\Delta E_{n,\mathbf{k}_c}=-\mathbf{B}\cdot\mathbf{m}_{n}(\mathbf{k}_c)$
where the orbital magnetic moment is given by 
\begin{equation}
\mathbf{m}_{n}(\mathbf{k}_c) =-\frac{e}{4m}\langle\Psi|(\hat{\mathbf{r}}-\mathbf{r}_{c})\times\hat{\mathbf{p}}|\Psi\rangle+\mathrm{h.c.},\label{eq:wavepacketM}
\end{equation}
where $\hat{\mathbf{p}}=\mathbf{p}\tau_{0}$
is the electron momentum operator acting on the particle-hole space.

The detailed calculation of this orbital magnetic moment is shown in the appendix. It resembles the one for the electron system~\cite{PRB1996Chang_electronOMM}. A straightforward integration over the wavepacket function gives,
\begin{equation}
\mathbf{m}_{n}(\mathbf{k}_{c})= 
 -\frac{e}{2m}{\rm Re}\left[\langle D_{\mathbf{k}}\phi_{n,\mathbf{k}}|\times\hat{\mathbf{p}}_{\mathbf{k}}|\phi_{n,\mathbf{k}}\rangle \right]\mid_{{\mathbf k}={\mathbf k}_c},\label{eq:OMM}
\end{equation}
where $D_{\mathbf{k}}\equiv i\partial_{\mathbf{k}}-\mathbf{A}_{\mathbf{k}}$ is the gauge invariant $\mathbf{k}$-space derivative~\cite{Arxiv2024liao_semiSC} with $\mathbf{A}_{\mathbf{k}}=i\langle\phi_{n,\mathbf{k}}|\partial_{\mathbf{k}}\phi_{n,\mathbf{k}}\rangle$ being the quasiparticle Berry connection, and ${\hat {\mathbf p}}_{\mathbf k} = e^{-i\mathbf{k}\cdot\mathbf{r}}\hat{\mathbf{p}}e^{i\mathbf{k}\cdot\mathbf{r}}$ is the $\mathbf{k}$-dependent momentum operator. From this formula we see that the orbital magnetic moment of each wavepacket depends only on its momentum-space center ${\mathbf k}_c$. Therefore, for simplicity, we omit the label $c$ in the momentum ${\mathbf k}_c$ below.

Now we try to explicitly evaluate the quantum averaging over the $\mathbf{k}$-dependent momentum operator ${\hat {\mathbf p}}_{\mathbf k}$ in Eq.~(\ref{eq:OMM}). For this purpose, we use the relation $\hat{\mathbf{p}}_{\mathbf{k}}= (m/\hbar)\tau_{z}\partial_{\mathbf{k}}\hat{H}^d_{\mathbf k}$, where $\hat{H}^d_{{\mathbf k}} ={\rm diag}(e^{-i\mathbf{k}\cdot\mathbf{r}}\hat h_0 e^{i\mathbf{k}\cdot\mathbf{r}}, -e^{-i\mathbf{k}\cdot\mathbf{r}}\hat h^*_0 e^{i\mathbf{k}\cdot\mathbf{r}})$ depends only on the diagonal blocks of the BdG Hamiltonian, and $\tau_{z}$ is the Pauli matrix acting on the particle-hole space. We plug this expression into Eq.~(\ref{eq:OMM}) and
insert the identity operator $\sum_{n^{\prime}}|\phi_{n^{\prime},\mathbf{k}}\rangle\langle\phi_{n^{\prime},\mathbf{k}}|$, and then the orbital magnetic moment can be written as 
\begin{align}
\mathbf{m}_{n}(\mathbf{k})&=\frac{e}{2\hbar}\sum_{n^{\prime}\neq n}\mathrm{Im}\left[\frac{\langle\phi_{n}|\partial_{\mathbf{k}}\hat{H}_{\mathbf{k}}|\phi_{n^{\prime}}\rangle\times\langle\phi_{n^{\prime}}|\tau_{z}\partial_{\mathbf{k}}\hat{H}^d_{\mathbf{k}}|\phi_{n}\rangle}{E_{n',\mathbf{k}}-E_{n,\mathbf{k}}}\right],\label{eq:OMMnumcal}
\end{align}
where $\hat{H}_{\mathbf{k}}=e^{-i\mathbf{k}\cdot\mathbf{r}}\hat{H}_{c}e^{i\mathbf{k}\cdot\mathbf{r}}$ is the $\mathbf{k}$-dependent Hamiltonian which satisfies the eigenvalue equation $\hat{H}_{\mathbf{k}}|\phi_{n,\mathbf{k}}\rangle=E_{n,\mathbf{k}}|\phi_{n,\mathbf{k}}\rangle$.

The expression (\ref{eq:OMMnumcal}) is the central result of this work. It is obviously gauge-invariant because the phase factors of eigen functions from the gauge transformation will cancel out. Comparing with the expression for Bloch electrons, there are two differences. Firstly, there is an extra $\tau_z$ factor that is coming from the effective charge of the BdG quasiparticle. Secondly, the formula involves not only the $\mathbf{k}$-dependent BdG Hamiltonian $H_{\mathbf k}$, but also its diagonal blocks $\hat{H}^d_{\mathbf k}$. This is due to the fact that the vector gauge does not enter the effective momentum of the pairing gap. As a result, the quasiparticle orbital magnetic moment cannot come solely from the $\mathbf{k}$-dependence of the gap function, which is entirely different from the behavior of the quasiparticle Berry curvatures~\cite{PRL2021Wang_semiSC,Arxiv2024liao_semiSC}.

For conventional $s$-wave superconductors where the pairing gap has no $\mathbf{k}$-dependence, we would have $\partial_{\mathbf{k}}\hat{H}_{\mathbf{k}}=\partial_{\mathbf{k}}\hat{H}^d_{\mathbf k}$. Then the quasiparticle orbital magnetic moment can be further expressed as derivatives to eigen functions,
\begin{align}
\mathbf{m}_{n}(\mathbf{k})= & \frac{e}{2\hbar}\mathrm{Re}\left[\langle D_{\mathbf{k}}\phi_{n,\mathbf{k}}|\times\tau_{z}(H_{\mathbf{k}}-E_{n,\mathbf{k}})|\partial_{\mathbf{k}}\phi_{n,\mathbf{k}}\rangle\right]\label{eq:OMMdphi}\\
 & -\frac{1}{2\hbar}\mathbf{d}_{e}\times\partial_{\mathbf{k}}E_{n,\mathbf{k}},\nonumber 
\end{align}
with $\rho_{\mathbf{k}}=e\langle\phi_{n,\mathbf{k}}|\tau_{z}|\phi_{n,\mathbf{k}}\rangle$ the effective charge of the wavepacket and $\mathbf{d}_{e}=e\textrm{Re}[\langle\phi_{n,\mathbf{k}}|\tau_{z}|i\partial_{\mathbf{k}}\phi_{n,\mathbf{k}}\rangle-\rho_{\mathbf{k}}\mathbf{A}_{\mathbf{k}}/e]$ the charge dipole of the wavepacket~\citep{Arxiv2024liao_semiSC}. The first term resembles the orbital magnetic moment of the electrons, but with an extra $\tau_z$ matrix in the quantum averaging. This originates from the effective charge of the wavepacket which is exactly the quantum averaging over a $\tau_z$ matrix. The second term is the dipole induced current which comes from the fact that the charge center of the wavepacket is different from the probability center of the wavepacket.

\section{\protect\label{sec:Quantum}Linear response theory for orbital magnetic moment in superconductors}
\enlargethispage{15\baselineskip}
We can verify the semiclassical result with a linear response approach that was developed in Ref.~\citep{PRL2007Shi_quantumOM} for studying the orbital magnetic moment of Bloch electrons. In this approach, the orbital magnetization can be obtained by the linear response of the grand thermodynamic potential at low temperature $K=E-\mu N$ with respect to the magnetic field. To get around the difficulty of the non-locality of the orbital magnetization operator, the magnetic field is introduced into the system with an infinitely slow spatial variation of
\begin{equation}
\mathbf{B}(\mathbf{r})=\frac{B}{2}(\cos q_{x}x+\cos q_{y}y)\hat{\mathbf{z}},\label{eq:MagneticField}
\end{equation}
where $B$ is the amplitude of the magnetic field, $q_x$ and $q_y$ are the wave vectors of the magnetic field along the two spatial directions. By adopting a convenient gauge, we can write down the corresponding vector potential as 
\begin{equation}\label{eq:vectorpotential}
\mathbf{A}(\mathbf{r})=\frac{B}{2}\left(\frac{\sin q_{x}x}{q_{x}}\hat{\mathbf{y}}-\frac{\sin q_{y}y}{q_{y}}\hat{\mathbf{x}}\right).
\end{equation}
We will calculate the linear response to this gauge field and then take the long-wavelength limit that $q_{x,y}$ goes to zero to obtain the orbital magnetic moment that couples with the static magnetic field. We note that in this long-wavelength limit our gauge choice corresponds to the conventional circular gauge which is the same gauge we took in semiclassical calculations.
Now the BdG Hamiltonian is written as
\begin{equation}
\hat{H}_{\rm{BdG}}=\left(\begin{array}{cc}
\hat h_0 [ \mathbf{ r},\mathbf{p}+e\mathbf{A}] & \hat \Delta(\mathbf{p})\\
\hat \Delta^{\dagger}(\mathbf{p}) & -\hat h^*_0 [ \mathbf{r},\mathbf{p}+e\mathbf{A}] 
\end{array}\right),
\end{equation}
where the vector potential $\mathbf{A}$ is given in Eq.~(\ref{eq:vectorpotential}). Now we can expand this BdG Hamiltonian to the linear order of the magnetic field ${\hat H}_{\rm{BdG}} = {\hat H}+ \Delta \hat{H}$ where 
${\hat H} = {\hat H}_{\rm{BdG}} ({\mathbf B} = 0)$
is the unperturbed BdG Hamiltonian with the quasiparticle band dispersion $E_{n,\mathbf{k}}$
and quasiparticle eigenstates $\psi_{n,\mathbf{k}}(\mathbf{r})=e^{i\mathbf{k}\cdot\mathbf{r}}\phi_{n,\mathbf{k}}(\mathbf{r})$, and $\Delta \hat{H}$ is the first-order perturbation Hamiltonian induced by the magnetic field that is written as
\begin{align}
\Delta \hat{H} & =\frac{e}{2m}\left[\mathbf{p}\cdot\mathbf{A}+\mathbf{A}\cdot\mathbf{p}\right]\tau_{0}.
\end{align}
This perturbed Hamiltonian will induce a correction to the grand canonical ensemble energy density~\cite{PRL2007Shi_quantumOM} $K(\mathbf{r})=\sum_{n,\mathbf{k}}f_{n,\mathbf{k}}\mathrm{Re}\left[\psi_{n,\mathbf{k}}^{*}(\mathbf{r}){\hat H}\psi_{n,\mathbf{k}}(\mathbf{r})\right]$ with $f_{n,\mathbf{k}}$ being the occupation number of quasiparticles. Then the orbital magnetization can be obtained
from the corresponding Fourier component $M_{z}=-\frac{2}{VB}\int\mathrm{d}\mathbf{r}\delta K(\mathbf{r})(\cos q_{y}y+\cos q_{x}x)$ where $V$ is the system volume.

The first-order correction to the grand canonical ensemble energy
density has contributions from the occupation
number, the Hamiltonian, and the wave function: 
\begin{align}
\delta K(\mathbf{r})= & \mathrm{Re}\sum_{n,\mathbf{k}}\left\{ \delta f_{n,\mathbf{k}}\psi_{n,\mathbf{k}}^{*}{\hat H}\psi_{n,\mathbf{k}}+f_{n,\mathbf{k}}\psi_{n,\mathbf{k}}^{*}\Delta\hat{H}\psi_{n,\mathbf{k}}\right.\nonumber \\
 & \left.+f_{n,\mathbf{k}}(\psi_{n,\mathbf{k}}^{*}{\hat H}\delta\psi_{n,\mathbf{k}}+\delta\psi_{n,\mathbf{k}}^{*}{\hat H}\psi_{n,\mathbf{k}})\right\} .\label{eq:deltaK}
\end{align}
The first two terms do not contribute to $M_{z}$ after performing
the spatial integration because of the cosine factors. The last two terms mainly involve the first-order corrections to the wave function which can be written by the standard first-order perturbation theory
\begin{widetext}
\begin{align} \label{eq:deltapsi}
\delta\psi_{n,\mathbf{k}}(\mathbf{r})= & \frac{eB}{8imq_{x}}\sum_{n^{\prime}}\left[\frac{e^{i(\mathbf{k}+\mathbf{q}_{x})\cdot\mathbf{r}}|\phi_{n^{\prime},\mathbf{k}+\mathbf{q}_{x}}\rangle\langle\phi_{n^{\prime},\mathbf{k}+\mathbf{q}_{x}}|\hat{p}_{\mathbf{k}}^{y}+\hat{p}_{\mathbf{k}+\mathbf{q}_{x}}^{y}|\phi_{n,\mathbf{k}}\rangle}{E_{n,\mathbf{k}}-E_{n^{\prime},\mathbf{k}+\mathbf{q}_{x}}}-(\mathbf{q}_{x}\rightarrow-\mathbf{q}_{x})\right]\\
 & -\frac{eB}{8imq_{y}}\sum_{n^{\prime}}\left[\frac{e^{i(\mathbf{k}+\mathbf{q}_{y})\cdot\mathbf{r}}|\phi_{n^{\prime},\mathbf{k}+\mathbf{q}_{y}}\rangle\langle\phi_{n^{\prime},\mathbf{k}+\mathbf{q}_{y}}|\hat{p}_{\mathbf{k}}^{x}+\hat{p}_{\mathbf{k}+\mathbf{q}_{y}}^{x}|\phi_{n,\mathbf{k}}\rangle}{E_{n,\mathbf{k}}-E_{n^{\prime},\mathbf{k}+\mathbf{q}_{y}}}-(\mathbf{q}_{y}\rightarrow-\mathbf{q}_{y})\right],\nonumber 
\end{align}
\end{widetext}
 where $\mathbf{q}_{x}=(q_{x},0,0)$ and $\mathbf{q}_{y}=(0,q_{y},0)$.
Substituting this wave function correction $\delta\psi_{n,\mathbf{k}}$ into Eq.~(\ref{eq:deltaK})
and performing the spatial integration of $M_z$, we can obtain the orbital magnetization $\bf M$ by taking the long-wavelength limit of $\mathbf{q}\rightarrow0$ and generalizing the result to other components~\cite{PRL2007Shi_quantumOM}. In the expression of the orbital magnetization, we find that significant contributions come from the orbital magnetic moment of the quasiparticles which is written as
\begin{align}
\mathbf{m}_{n}(\mathbf{k}) & =\frac{e}{2\hbar}\sum_{n^{\prime}\neq n}\mathrm{Im}\left[\frac{\langle\phi_{n}|\partial_{\mathbf{k}}\hat{H}_{\mathbf{k}}|\phi_{n^{\prime}}\rangle\times\langle\phi_{n^{\prime}}|\tau_{z}\partial_{\mathbf{k}}\hat{H}_{\mathbf{k}}^{d}|\phi_{n}\rangle}{E_{n^{\prime},\mathbf{k}}-E_{n,\mathbf{k}}}\right].\label{eq:OMMz}
\end{align}
Here we use the relation $\hat{\mathbf{p}}_{\mathbf{k}}= (m/\hbar)\tau_{z}\partial_{\mathbf{k}}\hat{H}^d_{\mathbf k}$ again. It is clear that this result is identical to that in Eq.~(\ref{eq:OMMnumcal}). This linear response approach for calculating the orbital magnetic moment verifies our semiclassical results.

\section{\label{sec:AngularMomentum}Orbital magnetic moment and orbital angular momentum in superconductors}

The effective charge of Bogoliubov quasiparticles is not a conserved quantity, because Bogoliubov quasiparticles are quantum superposition of electrons and holes where the superposition ratio may vary during dynamical processes. In the wavepacket approach, this is reflected in the fact that the charge center of the quasiparticle wavepacket generally does not coincide
with its probability center. Consequently, the expression for the orbital magnetic moment is quite unique for superconductors. We can see this more clearly by comparing the quasiparticle's orbital magnetic moment with its orbital angular momentum. 

The orbital angular momentum of a Bogoliubov quasiparticle can be obtained by the wavepacket averaging over the angular momentum operator $\hat {\mathbf L} = \hat {\mathbf r} \times \hat {\mathbf P}$ where $\hat {\mathbf r}$ and $\hat {\mathbf P}$ are the quasiparticle position and momentum operators. 
Following a similar derivation of Eq.~(\ref{eq:OMMnumcal}) and
taking the relation of $\hat{\mathbf{P}}_{\mathbf{k}}=(m/\hbar)\partial_{\mathbf{k}}\hat{H}_{\mathbf{k}}$
for the quasiparticle momentum operator, we obtain the angular momentum
of a wavepacket as 
\begin{equation}
\mathbf{L}_{n}(\mathbf{k})=-\frac{m}{\hbar}\sum_{n^{\prime}\neq n}\mathrm{Im}\left[\frac{\langle\phi_{n}|\partial_{\mathbf{k}}\hat{H}_{\mathbf{k}}|\phi_{n^{\prime}}\rangle\times\langle\phi_{n^{\prime}}|\partial_{\mathbf{k}}\hat{H}_{\mathbf{k}}|\phi_{n}\rangle}{E_{n',\mathbf{k}}-E_{n,\mathbf{k}}}\right].\label{eq:AngularMomentum}
\end{equation}
This expression for the orbital angular momentum formally resembles the one for electron systems and looks similar to the expression of the quasiparticle Berry curvature with the only difference in the denominator: the orbital angular momentum is the energy difference of two BdG bands while the Berry curvature is the square of the energy difference~\cite{PRL2021Wang_semiSC,Arxiv2024liao_semiSC}. Comparing Eq.~(\ref{eq:AngularMomentum}) and Eq.~(\ref{eq:OMMnumcal}), we can see that the difference between the orbital angular momentum and the orbital magnetic moment is the replacement of $\partial_{\mathbf{k}}\hat{H}_{\mathbf{k}}$ with $\tau_{z}\partial_{\mathbf{k}}\hat{H}_{\mathbf{k}}^{d}$ in the numerator. This difference is due to the fact that the self-rotation of the charge distribution is different from the self-rotation of the probability distribution within a wavepacket, which is a unique feature of the charge non-conserved Bogoliubov quasiparticles. The $\tau_z$ matrix in the expression of the magnetic moment comes from the superposition of the electron and hole components which have opposite charges, while the derivative over the diagonal component of the BdG Hamiltonian reflects the fact that the energy gap describes the relative wave function of the Cooper pair which does not directly couple with the gauge field.

The fact that the relative wave function of the energy gap does not directly couple with the orbital magnetic moment has qualitative impacts. As a result, a chiral superconductivity does not necessarily have nonzero orbital magnetic moment even though the orbital angular momentum is nonzero. To see this point clearly, we consider the simplest chiral $p$-wave superconductor which can be described by a $2\times2$ BdG Hamiltonian as,
\begin{equation}
\hat{H}_{\mathbf{k}}=\left(\begin{array}{cc}
\xi_\mathbf{k} & \Delta_0(k_x +ik_y)\\
\Delta_0(k_x - ik_y) & -\xi_{-\mathbf{k}}
\end{array}\right),
\end{equation}
where $\xi_\mathbf{k} = \hbar^{2}k^{2}/(2m)-\mu$ is the electron dispersion with $\mu$ the chemical potential and $\Delta_0$ represents the energy gap amplitude. For this model, the orbital angular momentum and the orbital magnetic moment are exactly solvable. We find a finite orbital angular momentum of $L(\mathbf{k}) =m\Delta_{0}^{2}(\xi_\mathbf{k}+2\mu)/(\hbar E^2_{\mathbf{k}})$ with $E_{\mathbf{k}}=\sqrt{\xi_{\mathbf{k}}^{2}+\Delta_{0}^{2}{k}^{2}}$ the quasiparticle dispersion, while the orbital magnetic moment vanishes in the entire momentum space
\begin{equation}
m(\mathbf{k})=0.
\end{equation}
The chiral
$p$-wave gap induces nonzero Berry curvature $\Omega(\mathbf{k})$ and orbital
angular momentum, and the two satisfy a relation $L(\mathbf{k})=-2(m/\hbar)E_{\mathbf{k}}\Omega(\mathbf{k})$ similar to that for Bloch electrons~\cite{PRL2007Xiao_Graphene}. In contrast, for orbital magnetic moment the derivative $\tau_z \partial_{\bf k} H^d_{\bf k}$ gives an identity matrix, therefore we have $\langle \phi_{n^{\prime}}| \tau_z \partial_{\bf k} H^d_{\bf k} |\phi_{n}\rangle = \partial_{\bf k} \xi_{\bf k} \langle \phi_{n^{\prime}}|\phi_{n}\rangle = 0$ since $n^{\prime}\neq n$. This behavior is quite different from Bloch electrons, where a finite Berry curvature is generically
accompanied by a finite orbital magnetic moment~\citep{PRB1996Chang_electronOMM,PRL2007Xiao_Graphene,RMP2010Xiao_BerryPhase}.

For a general two-band BdG Hamiltonian, a chiral superconducting gap can induce a nonzero orbital magnetic moment if the electron dispersion is not symmetric with respect to the momentum $\xi_{\bf k} \neq \xi_{\bf -k}$. Therefore, the diagonal part of the BdG Hamiltonian can be written as $\hat{H}_{\mathbf{k}}^{d}=\xi_{\mathbf{k}}^{a}\tau_{0}+\xi_{\mathbf{k}}^{b}\tau_{z}$, where $\xi_{\mathbf{k}}^{a} = ({\xi_{\mathbf{k}}-\xi_{-\mathbf{k}}})/{2}$ and $\xi_{\mathbf{k}}^{b} = ({\xi_{\mathbf{k}}+\xi_{-\mathbf{k}}})/{2}$. Then the quantum averaging writes as $\langle\phi_{n^{\prime}}|\tau_{z}\partial_{\mathbf{k}}\hat{H}_{\mathbf{k}}^{d}|\phi_{n}\rangle=\partial_{\mathbf{k}}\xi_{\mathbf{k}}^{a}\langle\phi_{n^{\prime}}|\tau_{z}|\phi_{n}\rangle$, where we take the orthogonality relation between the eigen function of different BdG
bands. The nonzero $\xi^{a}_{\bf k}$ can be achieved by breaking the spatial inversion symmetry or the time-reversal
symmetry in the electron Hamiltonian, or by driving a supercurrent
that produces a Doppler shift of the BdG spectrum~\cite{Sci2021Zhu_doppler,PRL2022Phan_doppler,Arxiv2024liao_semiSC}.
In multiband systems, additional internal structure in the electron Hamiltonian can make $\tau_{z}\partial_{\mathbf{k}}\hat{H}_{\mathbf{k}}^{d}$
not proportional to the identity operator, thereby yielding a nonzero
matrix element $\langle\phi_{n^{\prime}}|\tau_{z}\partial_{\mathbf{k}}\hat{H}_{\mathbf{k}}^{d}|\phi_{n}\rangle$,
as exemplified by the honeycomb lattice model discussed in Sec.~\ref{sec:ModelCalculation}.

\section{Responses induced by orbital magnetic moments}\label{sec:PhysicalProperties}

The orbital magnetic moment of Bogoliubov quasiparticles widely influences the spectroscopic and transport properties of superconducting systems. First of all, by definition the orbital magnetic moment modifies the energy spectrum with a term that is linear in the applied magnetic field. This energy correction induces a first-order energy
shift for the Bogoliubov band at momentum $\mathbf{k}$, 
\begin{equation}
\Delta E_{n,\mathbf{k}}=-\mathbf{B}\cdot\mathbf{m}_{n}(\mathbf{k}).\label{eq:deltaE}
\end{equation}
This energy shift for the quasiparticles can be observed by spectroscopic methods such as angle-resolved photoemission spectroscopy~\citep{RMP2003Damascelli_ARPES}. This should be a small energy shift because the magnetic field applied to superconductors usually cannot be too large. However, it may be large enough for experimental detection in the momentum areas where the orbital magnetic moment concentrates. Secondly, accompanying the shift of the quasiparticle energy spectrum, one expects a corresponding modification in the
density of states. For superconductors, the field-corrected density of states is given by 
\begin{equation} \label{eq:LDOS}
n(E)=\sum_{n}\int{d\mathbf{k}}\left[|u|^{2}\delta(E-E_{n,\mathbf{k}}^{\mathbf{m}})+|v|^{2}\delta(E+E_{n,\mathbf{k}}^{\mathbf{m}})\right],
\end{equation}
where $E_{n,\mathbf{k}}^{\mathbf{m}}=E_{n,\mathbf{k}}-\mathbf{B}\cdot\mathbf{m}_{n}(\mathbf{k})$ is the quasiparticle band dispersion corrected by the orbital magnetic moment, $u$ and $v$ are the electron and hole components of the BdG
eigen function. We compare the field-corrected
density of states $n(E)$ with the zero-field counterpart $n_{0}(E)=n(\mathbf{B}=0)$,
and define their difference as $\delta n(E)=n(E)-n_{0}(E)$.
 Due to the orbital magnetic moment, the presence of an external magnetic
field leads to a nonzero correction in the density of states. Because $\mathbf{m}_{n}(\mathbf{k})$
is $\mathbf{k}$-dependent, energy shifts $\Delta E_{n,\mathbf{k}}$
are generically nonuniform across the Brillouin zone and give rise to energy-dependent
enhancement and suppression in the local density of states, which can be measured directly via scanning tunneling spectroscopy~\citep{RMP2007Fischer_STS}. 

Aside from the influence on the spectroscopic properties, the quasiparticle orbital magnetic
moment can also give rise to orbital Hall transport in superconductors. Here we examine the orbital Nernst response where a temperature gradient is applied to the superconductor as a driving force for the quasiparticles. 
In the semiclassical picture, each wavepacket carries an orbital magnetic moment $\mathbf{m}_{{\bf k}_c}$ and moves with the velocity $\dot{\mathbf{r}}_{c}$~\cite{Arxiv2024liao_semiSC}. Combining together, a wavepacket contributes an
orbital magnetic moment current of $\mathbf{j}^\alpha={m}_{{\bf k}_c}^\alpha\dot{\mathbf{r}}_{c}$ where $\alpha=\{x,y,z\}$ denotes components of the orbital magnetic moment.
Summing over wavepackets gives the total orbital magnetic moment current density 
\begin{equation}
\mathbf{J}^{\alpha}(\mathbf{r})=\sum_{n}\int{d\mathbf{k}_c}f_{\mathbf{k}_{c}}{m}_{{\bf k}_c}^\alpha\dot{\mathbf{r}}_{c}|_{\mathbf{r}_{c}=\mathbf{r}}.
\end{equation}
The wavepacket velocity is determined by the equation of motions which has been derived for superconductors~\cite{PRL2021Wang_semiSC,Arxiv2024liao_semiSC}. For linear intrinsic Hall response, we can simplify the equation of motion as $\dot{\mathbf{r}}_c=-\dot{\mathbf{k}}_c\times\Omega_{\mathbf{k}_c} =(\nabla_{{\bf r}_c} E/\hbar) \times \Omega_{\mathbf{k}_c} $ and then the orbital magnetic moment current density writes as,
\begin{equation}
\mathbf{J}^{\alpha}(\mathbf{r})=\sum_{n}\frac{1}{\hbar}\int{d\mathbf{k}_c}f_{\mathbf{k}_c} {m}_{{\bf k}_c}^\alpha \left[ \nabla_{{\bf r}_c} E \times\Omega_{\mathbf{k}_c} \right]_{\mathbf{r}_{c}=\mathbf{r}},
\end{equation}
where $f_{\mathbf{k}_c}$ reduces to the Fermi distribution function because we take the zeroth-order approximation for the occupation number of the quasiparticles. Now we take care of the spatial gradient to the energy $\nabla_{{\bf r}_c} E$ in the presence of a temperature gradient by introducing the quasiparticle grand potential function $g(E,T)=k_{{\rm B}}T\ln(1-f)$ with $k_{{\rm B}}$ the Boltzmann constant and $T$ the temperature. It satisfies the relation $f=\partial g/\partial E$ and the transport part of the orbital magnetic moment current can be expressed as $\mathbf{J}^{\alpha}(\mathbf{r}) = {\boldsymbol{\eta}}^\alpha \times\nabla_{\bf r} T $ with the orbital Nernst coefficient given by 
\begin{align} \label{eq:ONE}
{\boldsymbol{\eta}}^\alpha= \sum_{n}\frac{1}{\hbar}\int d\mathbf{k}_c\frac{\partial g}{\partial T}{m}_{{\bf k}_c}^\alpha {\boldsymbol \Omega}_{\mathbf{k}_c}. 
\end{align}
This formula resembles the one for the spin Nernst coefficient of the superconductors only with the effective spin of each wavepacket replaced by the orbital magnetic moment~\cite{PRB2021Xiao_Conserved,Arxiv2024liao_semiSC}. It requires the simultaneous existence of the orbital magnetic moment and the Berry curvature of Bogoliubov quasiparticles, which places a stricter constraint on the structure of the BdG Hamiltonian.

We finally comment on the robustness of the physical consequences discussed above. The magnetic-field-induced quasiparticle energy shift and the associated density of states correction are expected to be relatively robust, because they originate from the linear-in-field correction to the BdG quasiparticle energy. This situation is analogous to the case of Bloch electrons, where semiclassical theory shows that the orbital magnetic moment modifies the band energy in a magnetic field, and related energy shifts have been observed in valley-splitting measurements and attributed to the coupling between the magnetic field and the orbital magnetic moment~\cite{PRL2020Lee_Zeeman,NC2020Moriya_splitting}. In the present work, the quasiparticle orbital magnetic moment is defined in the same physical sense. By contrast, the orbital Nernst effect should be interpreted more cautiously, since existing studies of orbital Nernst transport have used different theoretical approaches and definitions of the orbital current~\cite{PRM2022Salemi_ONE,Nanolett2024Go_ONE,PRB2025An_ONE,PRB2025Ji_ONE,NPJ2026Markus_ONE}, and orbital Hall-type responses may also receive additional quantum corrections~\cite{PRL2025Liu_OHE}. Thus, the orbital Nernst coefficient derived here should be viewed as a semiclassical wavepacket formulation of orbital magnetic moment transport in superconducting quasiparticles.

\section{\protect\label{sec:ModelCalculation}Model calculations}

We have demonstrated that generating a quasiparticle orbital magnetic moment requires a suitable combination of electronic band structure and superconducting pairing function.
Here we consider a two-dimensional tight-binding model with a honeycomb
lattice that was introduced in Ref.~[\onlinecite{PRB2008Jiang_GrapheneModel}].
This model inherently possesses the sub-lattice degrees of freedom, which, combined with a chiral pairing gap, fulfills the symmetry and structural requirements for generating a finite orbital magnetic moment. The system consists of two sub-lattices with the Hamiltonian given by
\begin{eqnarray}
\hat{H}= & - & t \sum_{i,\beta,\sigma}[c_{A,{i},\sigma}^{\dagger}c_{B,{i}+\mathbf{\delta}_{\beta},\sigma}+\mathrm{h.c.}] \\
 & - & \mu \sum_{i,\sigma}(c_{A,{i},\sigma}^{\dagger}c_{A,{i},\sigma}+c_{B,{i}+\mathbf{\delta}_{1},\sigma}^{\dagger}c_{B,{i}+\mathbf{\delta}_{1},\sigma})\nonumber \\
 & + & \sum_{i,\beta}[\Delta_{\mathbf{\delta}_{\beta}}(c_{A,{i},\uparrow}^{\dagger}c_{B,{i}+\mathbf{\delta}_{\beta},\downarrow}^{\dagger}-c_{A,{i},\downarrow}^{\dagger}c_{B,{i}+\mathbf{\delta}_{\beta},\uparrow}^{\dagger})+\mathrm{h.c.}]\nonumber,
\end{eqnarray}
where $A$ and $B$ denote the two sub-lattices of a unit cell, $\sigma=\uparrow,~\downarrow$ denotes the
spin index, ${i}$ denotes the $A$ sub-lattice site with position $\mathbf{r}_{i}$, $\mathbf{\delta}_{\beta}$ with $\beta = \{1,2,3\}$ denotes the three vectors that connect the nearest $A$ and $B$ sub-lattices which are given as $\mathbf{\mathbf{\delta}}_{1}=(a,-\sqrt{3}a)/(2\sqrt{3})$, $\mathbf{\mathbf{\delta}}_{2}=(a,\sqrt{3}a)/(2\sqrt{3})$ and $\mathbf{\mathbf{\delta}}_{3}=(-a,0)/\sqrt{3}$ with $a$ the lattice constant, $c_{A}^{\dagger}$ and $c_{B}^{\dagger}$ are the creation operators on the $A$ and $B$ sub-lattices respectively, $\Delta_{\mathbf{\delta}_{\beta}}=\Delta e^{2i\beta \pi /3}$ represents the chiral $d$-wave pairing in the honeycomb lattice with $\Delta$ the pairing amplitude, and $t$ is 
the nearest-neighbor hopping amplitude.

This Hamiltonian can be transformed to the momentum space through the Fourier transformation of $c_{A,\mathbf{k},\sigma}^{\dagger}=(1/\sqrt{N})\sum_{i}e^{i\mathbf{k}\cdot\mathbf{r}_{i}}c_{A,i,\sigma}^{\dagger}$ and $c_{B,\mathbf{k},\sigma}^{\dagger}=(1/\sqrt{N})\sum_{i}e^{i\mathbf{k}\cdot(\mathbf{r}_{i}+\mathbf{\delta}_{\beta})}c_{B,i+\mathbf{\delta}_{\beta},\sigma}^{\dagger}$, where $N$ is the number of sites of each sub-lattice. By defining a Nambu spinor of $\Psi_{{\bf k}}=(c_{A,{\bf k},\uparrow},c_{B,{\bf k},\uparrow},c_{A,-{\bf k},\downarrow}^{\dagger},c_{B,-{\bf k},\downarrow}^{\dagger})^{T}$, the Hamiltonian can be compactly written as $\hat{H}=\sum_{{\bf k}}\Psi_{{\bf k}}^{\dagger}\hat{H}_{{\bf k}}\Psi_{{\bf k}}$ where we obtain a $4\times4$ BdG matrix
\begin{equation} \label{eq:modelH}
\hat{H}_{{\bf k}}=\left(\begin{array}{cccc}
-\mu & \tilde{f}_{\mathbf{k}} & 0 & \Delta_{\mathbf{k}}\\
\tilde{f}_{\mathbf{k}}^{*} & -\mu & \Delta_{-\mathbf{k}} & 0\\
0 & \Delta_{-\mathbf{k}}^{*} & \mu & -\tilde{f}_{-\mathbf{k}}^{*}\\
\Delta_{\mathbf{k}}^{*} & 0 & -\tilde{f}_{-\mathbf{k}} & \mu
\end{array}\right),
\end{equation}
with $\tilde{f}_{\mathbf{k}}=-t\sum_{\beta =1}^{3}e^{i\mathbf{k}\cdot\mathbf{\delta}_{\beta}}$ and $\Delta_{\mathbf{k}}=\sum_{\beta=1}^{3}\Delta_{\mathbf{\mathbf{\delta}}_{\beta}}e^{i\mathbf{k}\cdot\mathbf{\mathbf{\delta}}_{\beta}}$. This Hamiltonian provides a simple example for the numerical calculation of the orbital magnetic moments.

\begin{figure}[t]
\begin{centering}
\includegraphics[width=3.4in]{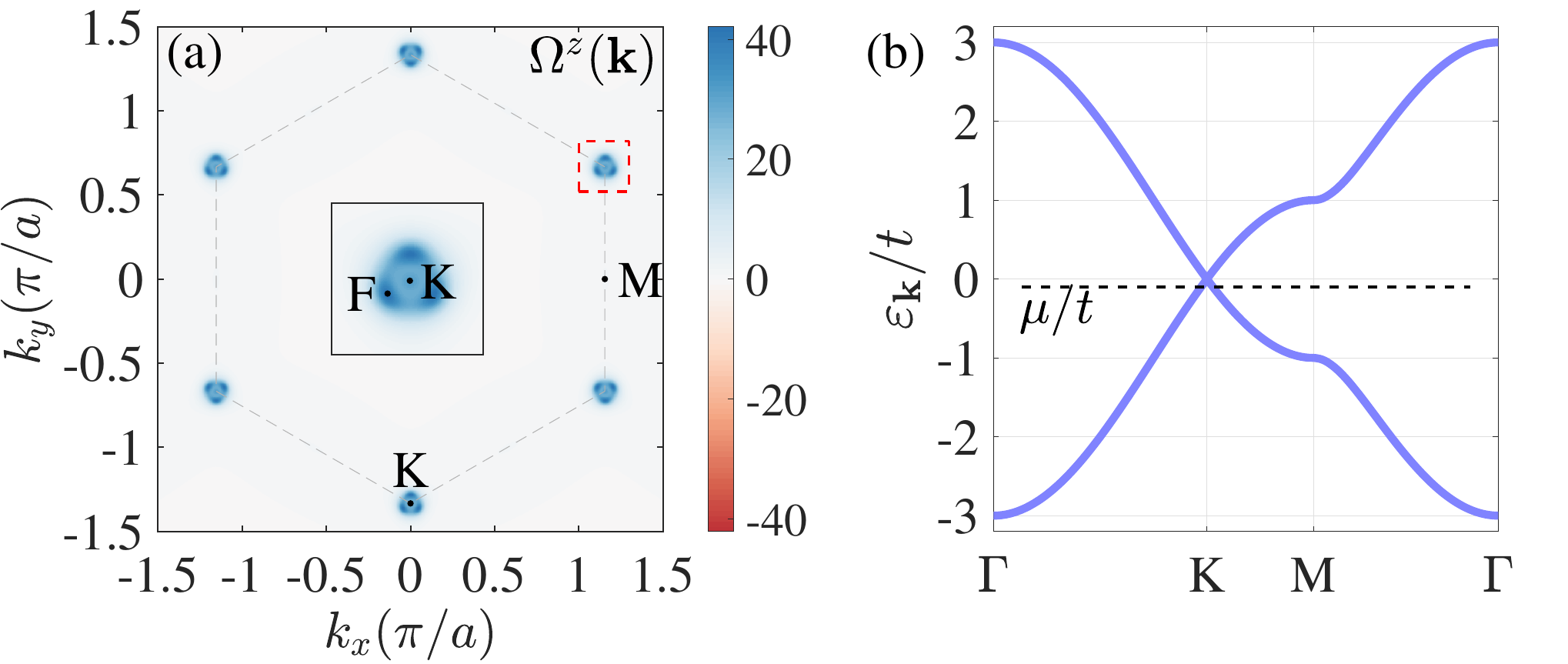}
\par\end{centering}
\begin{centering}
\includegraphics[width=3.4in]{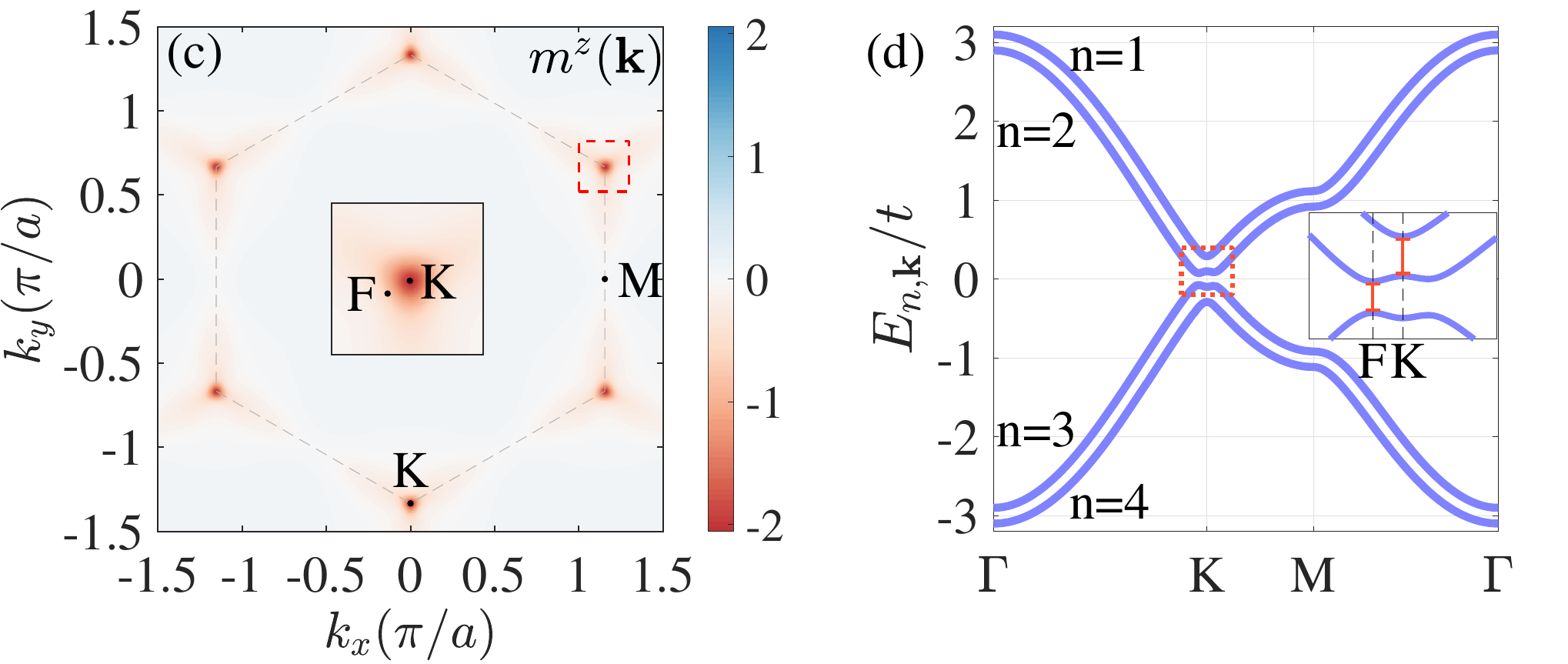}
\par\end{centering}
\begin{centering}
\includegraphics[width=3.4in]{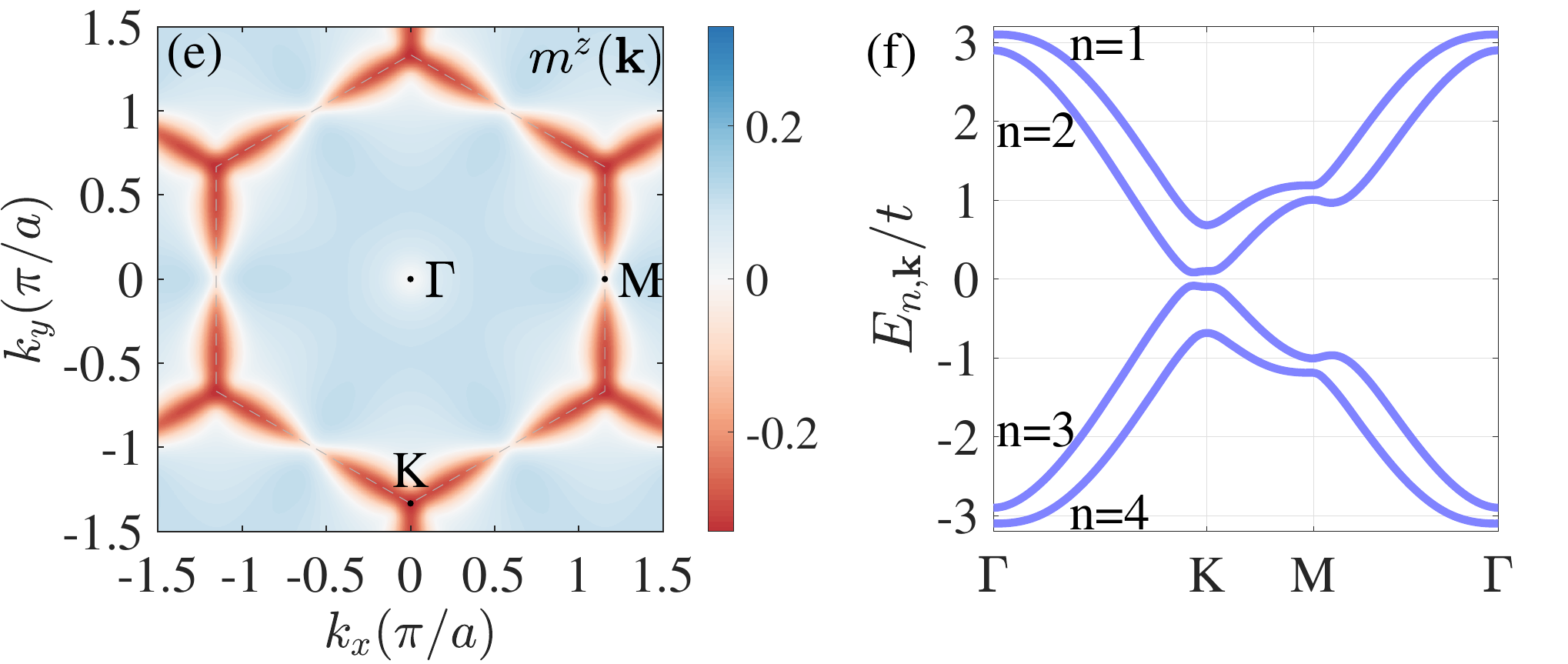}
\par\end{centering}
\caption{\protect(a) Momentum-space distribution of the Berry curvature $\Omega^{z}(\mathbf{k})$
(in units of $a^{2}$) for the lowest positive-energy band $n=2$. The gray dashed lines denote the boundary of the first Brillouin zone. The high-symmetry points are marked with black dots. The inset is the zoom-in view of the red-dashed region. The parameters are taken as $\mu/t=-0.1$ and $\Delta/t$=0.09. (b) Electronic band structure along
the high-symmetry lines. The dashed line indicates the chemical potential with $\mu/t=-0.1$. (c)
Momentum-space distribution of the orbital magnetic moment $m^{z}(\mathbf{k})$
(in units of $a^{2}et/\hbar$) for the lowest positive-energy band $n=2$, with
the parameters taken the same as in panel~(a). (d) Quasiparticle band structure
along the high-symmetry lines with parameters taken the same as in panel~(a). (e) and (f) The momentum-space distribution of the orbital magnetic moment and the quasiparticle band structure
along the high-symmetry lines with a different parameter of $\Delta/t$=0.225. The chemical potential is taken the same as in panel~(a).}\label{fig:k_space_distribution}
\end{figure}

\subsection{Orbital magnetic moment distribution in momentum space}
The electron band structure of the honeycomb-lattice system has Dirac cones
at the $K$ points of the Brillouin zone as shown in Fig.~\ref{fig:k_space_distribution}(b).
In order for a robust superconducting gap, a finite area of electron
Fermi surface is required. For this purpose, we consider the system to be tuned to deviate slightly from charge neutrality with the chemical potential 
shown as the dashed line in Fig.~\ref{fig:k_space_distribution}(b).
In this case, the electron Fermi surface is shaped as six small pockets
around the six Dirac points. The quasiparticle Berry curvature of
this model has been calculated~\citep{PRL2021Wang_semiSC,Arxiv2024liao_semiSC}.
As shown in Fig.~\ref{fig:k_space_distribution}(a), the quasiparticle
Berry curvature is induced by the chiral $d$-wave pairing and concentrates
on the electron Fermi surfaces.

Now we calculate the distribution of orbital magnetic moments in the
momentum space for this model. In Fig.~\ref{fig:k_space_distribution}(c), we
show the orbital magnetic moment $m^{z}(\mathbf{k})$ computed for the lowest positive-energy band
under the same parameters as in Fig.~\ref{fig:k_space_distribution}(a).
Although the orbital magnetic
moment is also predominantly concentrated near the $K$ points similar to the behavior of Berry curvatures, a clear distinction emerges between the fine structures of their distributions. This can be clearly seen from the insets of Figs.~\ref{fig:k_space_distribution}(a) and \ref{fig:k_space_distribution}(c). The Berry
curvature is peaked at the $F$ points where the electronic
Fermi surface resides, while the orbital magnetic moment peaks exactly at the $K$ point. This is quite different from Bloch electrons where the peaks of the orbital magnetic moment
and Berry curvatures often coincide with each other~\cite{PRL2007Xiao_Graphene}.

To understand this contrasting behavior, we analyze the BdG band structure
shown in Fig.~\ref{fig:k_space_distribution}(d). Both the orbital
magnetic moment and the Berry curvature depend on the energy differences between BdG bands. Specifically, the orbital magnetic moment scales as $1/(E_{n^{\prime},\mathbf{k}}-E_{n,\mathbf{k}})$, whereas
the Berry curvature scales as $1/(E_{n^{\prime} ,\mathbf{k}}-E_{n,\mathbf{k}})^{2}$~\cite{PRL2021Wang_semiSC,Arxiv2024liao_semiSC}. Thus,
small energy gaps may enhance both quantities. Focusing on the band $n=2$ shown in Fig.~\ref{fig:k_space_distribution}(d), we observe that the global minimum of the energy difference with $n'=3$ happens at $F$ points. This small band gap is responsible for the large quasiparticle Berry curvatures as shown in Fig.~\ref{fig:k_space_distribution}(a). However, the orbital magnetic moment is suppressed, because the matrix element $\langle\phi_{n^{\prime},\mathbf{k}}|\tau_{z}\partial_{\mathbf{k}}\hat{H}_{\mathbf{k}}^{d}|\phi_{n,\mathbf{k}}\rangle$
for bands $n=2$ and $n^\prime=3$ vanishes at the $F$ points. The other local minimum of the energy difference happens at $K$ points between the two bands $n=2$ and $n'=1$. This local band gap minimum is responsible for the enlargement of the orbital magnetic moment since the 
matrix elements $\langle\phi_{n^{\prime},\mathbf{k}}|\tau_{z}\partial_{\mathbf{k}}\hat{H}_{\mathbf{k}}^{d}|\phi_{n,\mathbf{k}}\rangle$
for bands $n=2$ and $n^\prime=1$ are nonzero.

In Fig.~\ref{fig:k_space_distribution}(e), we show the momentum-space distribution of the orbital magnetic
moment for a larger superconducting gap. In this case, the orbital magnetic moment
is no longer strongly localized at the $K$ points but becomes
more extended across the Brillouin zone. The corresponding band structure, as shown in Fig.~\ref{fig:k_space_distribution}(f), reveals that 
the overall energy gaps between different bands are generally larger
than those in Fig.~\ref{fig:k_space_distribution}(d), therefore, the magnitudes of the orbital magnetic
moment are smaller than the latter.

\subsection{Spectroscopic properties: energy correction and density of states correction}
\begin{figure}[t]
\begin{centering}
\includegraphics[width=3.4in]{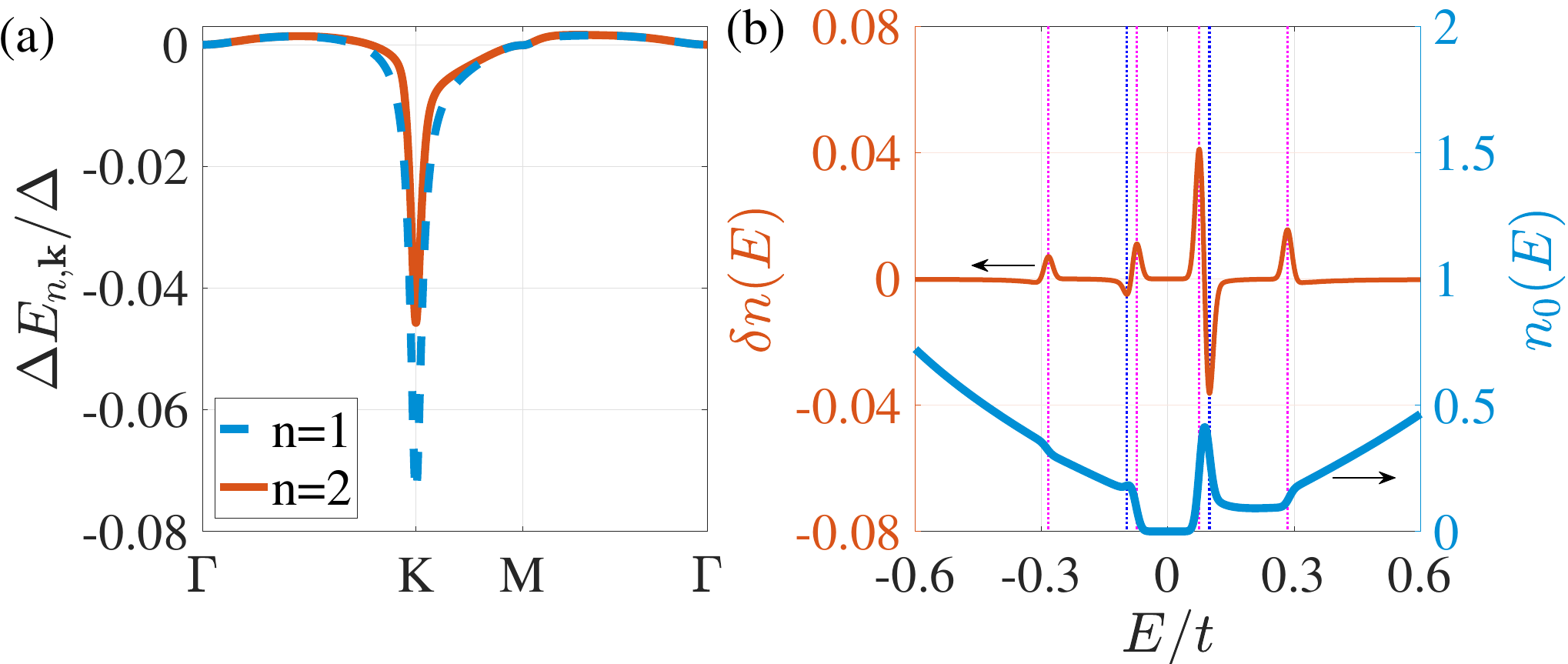} 
\par\end{centering}
\begin{centering}
\includegraphics[width=3.4in]{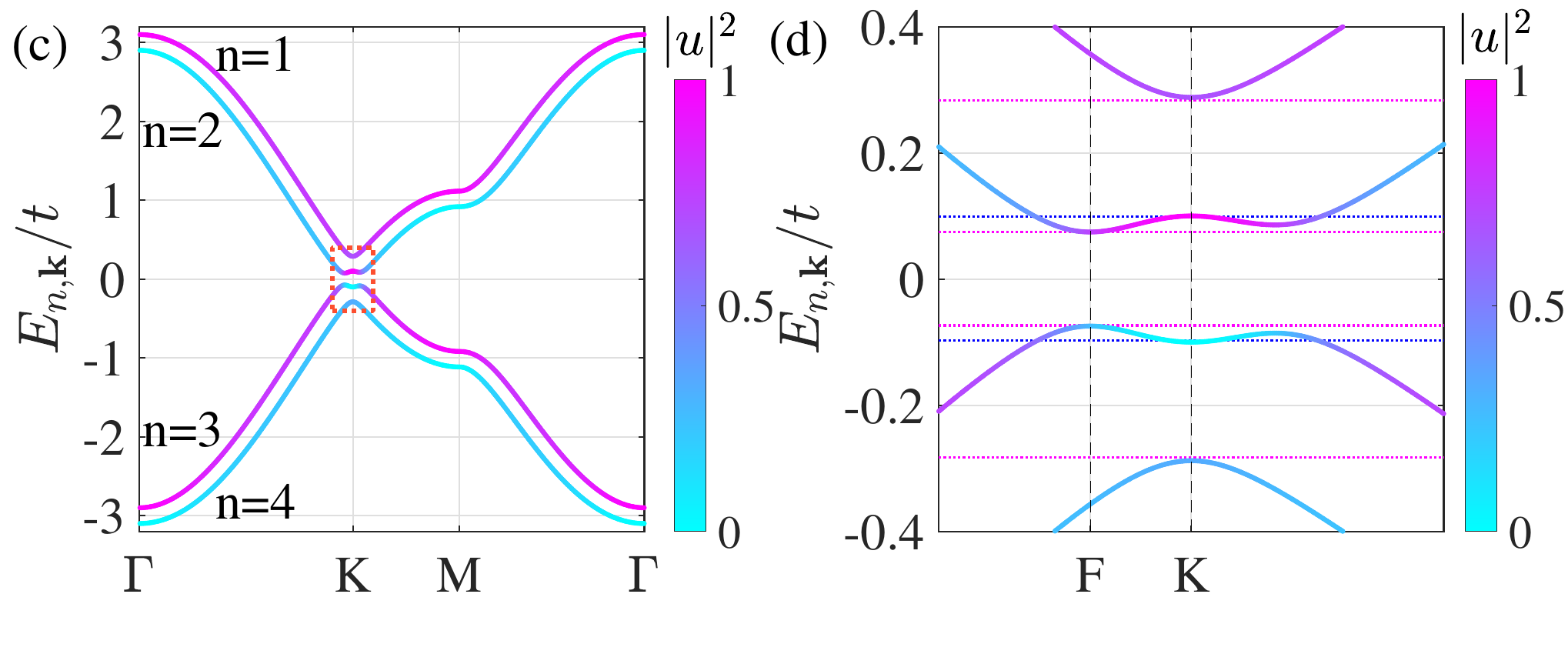} 
\par\end{centering}
\begin{centering}
\includegraphics[width=3.4in]{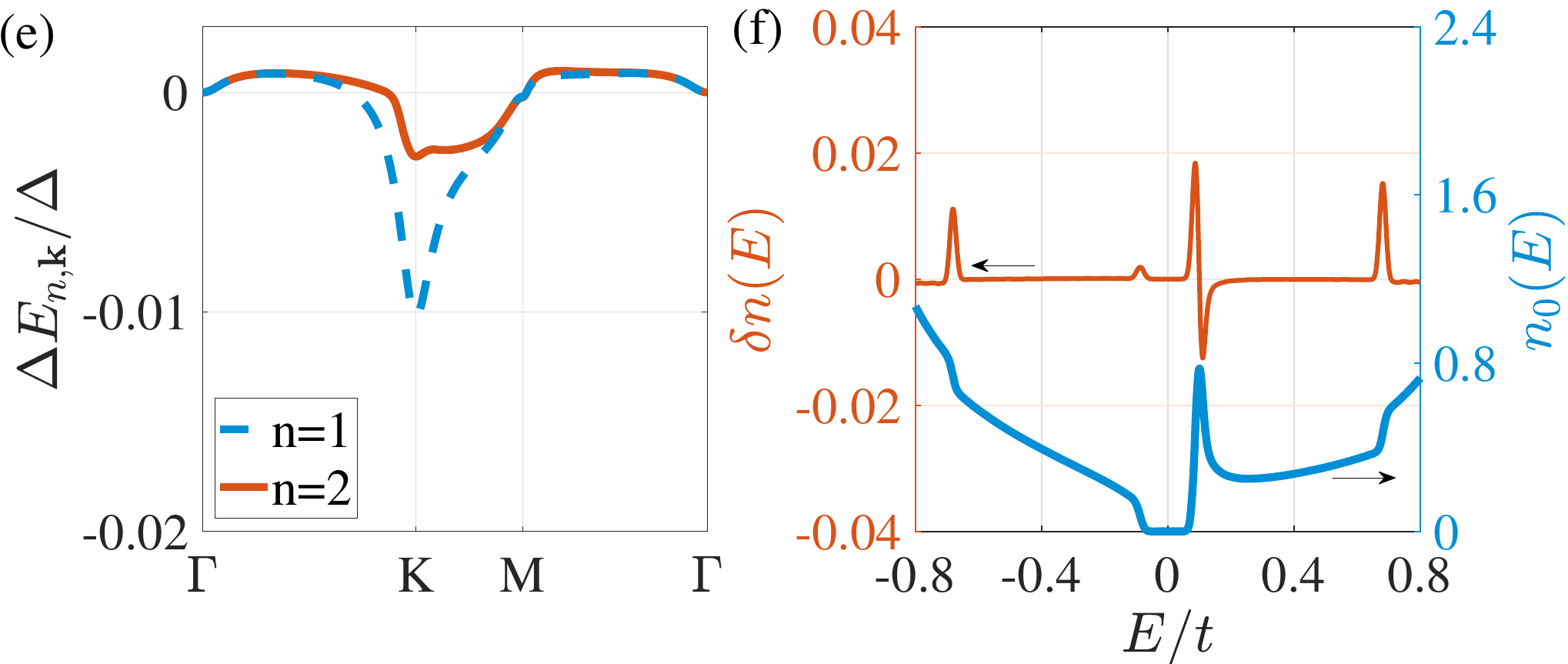} 
\par\end{centering}
\caption{\protect\label{fig:dE_and_DOS}(a) The ratio between energy correction
$\Delta E_{n\mathbf{k}}$ due to a uniform magnetic field and pairing
amplitude $\Delta$ for the positive-energy quasiparticle bands. (b) Field-induced density of states correction $\delta n(E)$ (orange line, left axis) and local density of states $n_{0}(E)$ in the absence of a magnetic field (blue line, right axis).
The peaks and dips in $\delta n(E)$ are marked by pink dashed lines and blue dashed lines, respectively.
(c) Quasiparticle band structure. The color scale represents the weight of the electron component $|u|^2$ with respect to the local density of states. (d) Zoom-in view of the red-dashed region in panel~(c), showing the detailed band structure near the $K$ point and $F$ point. The horizontal dashed lines mark the characteristic energies corresponding to the peaks and dips in the density of states correction $\delta n(E)$ shown in panel~(b), using the same color coding. (e) Same as panel (a), but for a larger pairing amplitude. (f) Same as panel (b), but for a larger pairing amplitude. Panels (a)-(d) are calculated with $\Delta/t=0.09$, while panels (e) and (f) are calculated with $\Delta/t=0.225$. The other parameters are $B_{z}=-(\hbar/a^2e)/500$ and $\mu/t=-0.1$.
}
\end{figure}

The distribution of the orbital
magnetic moment in momentum space directly influences the momentum-space energy spectrum in the presence of a magnetic field, as shown in Eq.~(\ref{eq:deltaE}). Here we perform numerical calculations of this energy correction for the honeycomb lattice with a 
uniform magnetic field $\mathbf{B}=B_{z}\hat{z}$. We present the energy correction $\Delta E_{n,\mathbf{k}}$
for the two positive energy BdG bands in Fig.~\ref{fig:dE_and_DOS}(a), while the corrections to the negative-energy BdG bands can be obtained through particle-hole
symmetry. It is obvious that the shift of the energy spectrum by the magnetic field coincides with the distribution of the orbital magnetic moment. 
The significant
energy shifts concentrate near the $K$
points in the momentum space, predominantly reducing the quasiparticle energies,
and the energy correction of band $n=1$ is larger than those of band $n=2$. As a result, the BdG gaps are reduced by the energy shift from the coupling of orbital magnetic moment with the external field. 

The corrections to the energy spectrum will modulate the density of states as shown in Eq.~(\ref{eq:LDOS}). We present the zero-field density of states $n_{0}(E)$
and its correction $\delta n(E)$ in Fig.~\ref{fig:dE_and_DOS}(b). The $\delta n(E)$ curves exhibit distinct peaks and dips, indicating energies
where the orbital magnetic moment has the largest modifications to
the density of states. Notably, the $\delta n(E)$ curve shows an
equal number of peaks and dips in both $E>0$ and $E<0$ regions,
occurring at symmetric energies. In Figs.~\ref{fig:dE_and_DOS}(c)
and \ref{fig:dE_and_DOS}(d), we present the BdG band structure and mark these characteristic energies. We find that the two peaks
closest to $E=0$ correspond to the energies at the $F$ points
of bands $n=2$ and $n=3$. As shown in Fig.~\ref{fig:dE_and_DOS}(a),
the orbital magnetic moment reduces the energies near the $K$ points
of the positive-energy bands, including those at the $F$ points of the 
band $n=2$. It also increases the energies at the $F$ points of
the band $n=3$, effectively narrowing the superconducting gap.
This increases the number of in-gap states and results in a positive density of states correction which generates the two
peaks nearest to $E=0$ in $\delta n(E)$. The two dips correspond
to the energies at the $K$ points of bands $n=2$ and $n=3$. The
substantial orbital magnetic moment at these points induces significant
energy shifts, therefore, both the energy decreases at the $K$ points of the band
$n=2$ and the energy increases at the $K$ points of the band $n=3$ 
reduce the number of states and lead to the observed dips in $\delta n(E)$. The two peaks farthest
from $E=0$ originate from energy modifications at the $K$ points
of bands $n=1$ and $n=4$. The large orbital magnetic moments at
these points cause the energy of band $n=1$ to decrease and that
of band $n=4$ to increase. These shifts enhance the number of states
within the corrected energy intervals, thereby forming the peak
structures in $\delta n(E)$.

Furthermore, we observe that the peaks and dips
in $\delta n(E)$ have different magnitudes in the $E>0$ and $E<0$ regions,
with a larger magnitude in the former region. This asymmetry
can be understood by examining the contributions from the electron
and the hole components of the quasiparticle wave functions. The local
density of states is determined by both the quasiparticle energy $E_{n,\mathbf{k}}$ and the squared
moduli of the electron and the hole wave functions, $|u|^{2}$ and $|v|^{2}$,
as given in Eq.~(\ref{eq:LDOS}). We show the value of $|u|^{2}$ as the color map in Fig.~\ref{fig:dE_and_DOS}(c) and Fig.~\ref{fig:dE_and_DOS}(d), and the value of $|v|^{2}$ can be obtained by the normalization requirement $|v|^{2}=1-|u|^{2}$.
We find that around the $F$ points the positive-energy $n=2$ band has an overwhelming $|u|^{2}$ weight while the negative-energy $n=3$ band has an overwhelming $|v|^{2}$ weight. As a result, the weight factors
determining the density of states of the energy at the positive-energy peak are
significantly larger than those at the corresponding negative-energy
peak. Therefore, it is natural that the density of states correction is larger in the positive-energy region. 

We also analyze the spectroscopic properties at a larger pairing amplitude, as shown in Figs.~\ref{fig:dE_and_DOS}(e) and \ref{fig:dE_and_DOS}(f). For $\Delta/t=0.225$, the orbital magnetic moment is smaller in magnitude than that for $\Delta/t=0.09$, as can be seen by comparing Figs.~\ref{fig:k_space_distribution}(c) and \ref{fig:k_space_distribution}(e). Consistently, the field-induced energy correction is also strongly reduced. The largest correction still occurs near the $K$ point, but its profile becomes more extended along the high-symmetry line, following the broader momentum-space distribution of the orbital magnetic moment in Fig.~\ref{fig:k_space_distribution}(e).
The corresponding density of states correction is shown in Fig.~\ref{fig:dE_and_DOS}(f). Similar to the case of $\Delta/t=0.09$, $\delta n(E)$ still exhibits several peak and dip structures. These structures appear at similar types of characteristic energies, namely near the superconducting gap edges and near the energies associated with the $K$ point in the quasiparticle spectrum. The main difference is that the overall magnitude of $\delta n(E)$ is smaller, reflecting the reduced orbital magnetic moment. In addition, the dip on the negative-energy side, which is clearly visible for $\Delta/t=0.09$, becomes much less pronounced for $\Delta/t=0.225$. This suppression comes from both the smaller orbital magnetic moment and the relatively weak weight of the density of states in the negative-energy region.

\begin{figure}[t]
\begin{centering}
\includegraphics[width=3.42in]{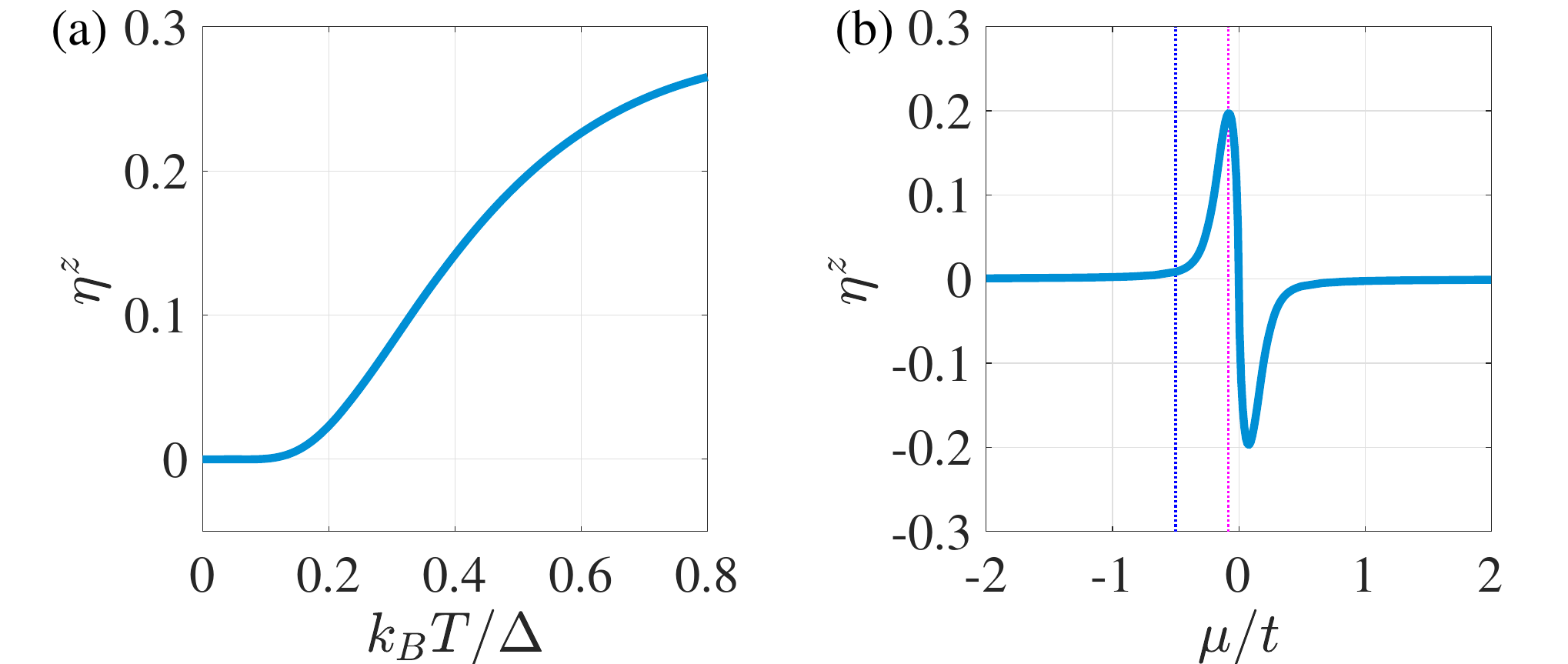}
\par\end{centering}
\begin{centering}
\includegraphics[width=3.4in]{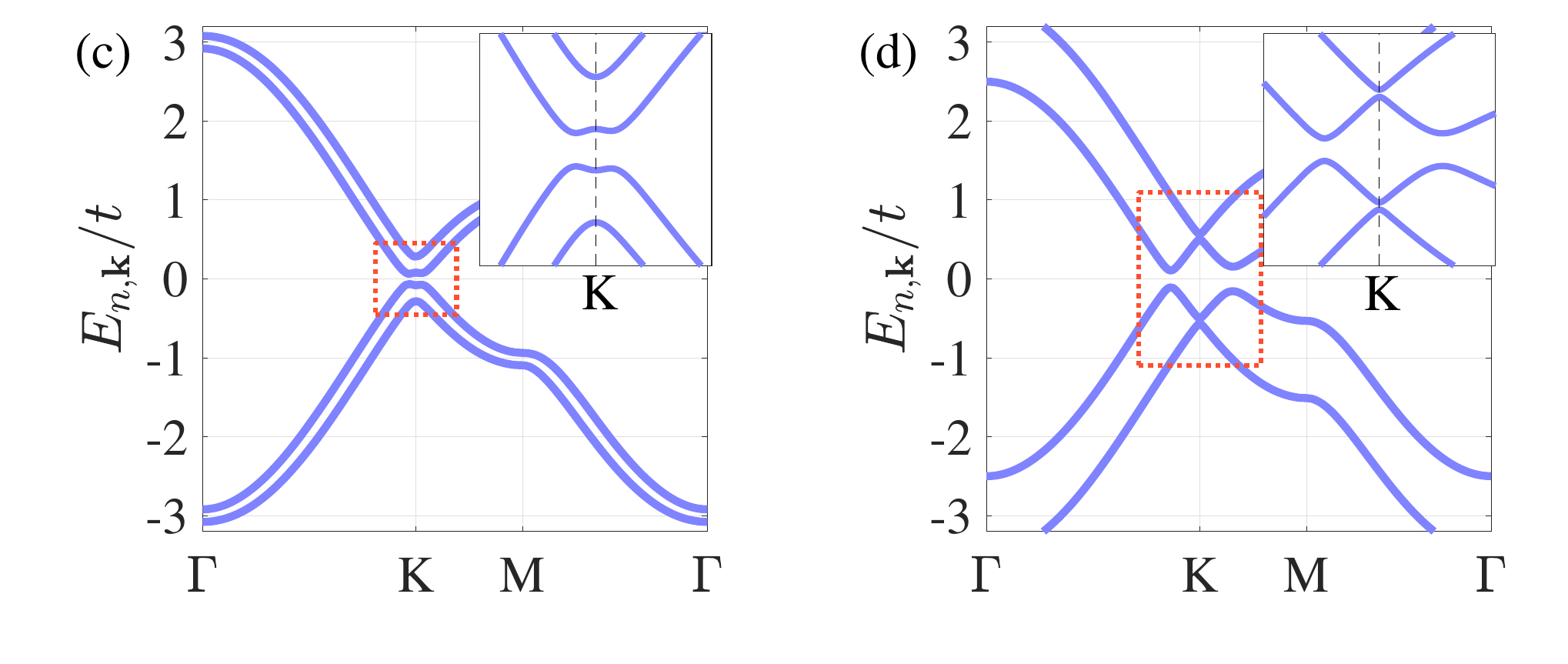}
\par\end{centering}
\caption{\protect\label{fig:Nernst}(a) Temperature dependence of the orbital
Nernst coefficient $\eta^{z}$ (in units of $a^{2}etk_B/\hbar^2$) at fixed $\mu/t=-0.1$. (b) Chemical potential dependence of the orbital
Nernst coefficient at fixed $k_{B}T/\Delta=0.5$. (c) Quasiparticle band structure
along the high-symmetry lines with $\mu/t=-0.08$. This chemical potential corresponds to the position marked by the pink dashed line in panel~(b). The inset shows a detailed view near the $K$
point. (d) Quasiparticle band structure with $\mu/t=-0.5$. This chemical potential corresponds to the position marked by the blue dashed line in panel~(b). The other parameter used is $\Delta/t=0.09$.}
\end{figure}

\subsection{Transport property: orbital Nernst effect}
The orbital magnetic moment not only strongly influences the spectroscopic properties of superconductors, but also induces the orbital Nernst effect in combination with non-trivial Berry curvatures as shown in Eq.~(\ref{eq:ONE}). For the two-dimensional model Hamiltonian Eq.~(\ref{eq:modelH}), the orbital Nernst coefficient ${\boldsymbol \eta}^\alpha$ becomes a scalar ${ \eta}^\alpha$.
We calculate $\eta^{z}$ and show the temperature and chemical
potential dependence in Fig.~\ref{fig:Nernst}(a) and Fig.~\ref{fig:Nernst}(b). The temperature dependence of the orbital Nernst coefficient $\eta^{z}$ is quite simple: it increases monotonically with increasing temperature. This is due to the increase of the number of quasiparticles which is roughly determined by the factor of $1/(1+e^{\Delta/k_B T})$. As for the chemical
potential dependence, we find that $\eta^{z}$ is exactly zero when the chemical potential locates at the Dirac point, which may simply be due to the vanishing electron density of states. When the chemical potential deviates from the Dirac points, two peaks appear and then $\eta^{z}$ gradually decreases.

To understand these transport features, we show the BdG band structure for two typical chemical potentials in Fig.~\ref{fig:Nernst}(c) and Fig.~\ref{fig:Nernst}(d), focusing on the $K$ point where a substantial orbital magnetic moment emerges.
The band structure shown in Fig.~\ref{fig:Nernst}(c) corresponds to the chemical potential at the peak of the orbital Nernst coefficient marked by the pink dashed line in Fig.~\ref{fig:Nernst}(b). It exhibits small band gaps at the $K$ points which generate large orbital magnetic moments and consequently give rise to the significant orbital Nernst effect. As the chemical potential shifts away from the Dirac point to the position indicated by the blue dashed line in Fig.~\ref{fig:Nernst}(b), the orbital Nernst coefficient diminishes. The corresponding band structure is shown in Fig.~\ref{fig:Nernst}(d). We can see that while the band gaps at the $K$ points remain small, the energies of the $K$ points now deviate notably from zero, significantly suppressing the magnitude of the derivative $\partial g/\partial T$. Consequently, despite the persistent small band gap, the orbital Nernst coefficient becomes smaller.

\section{\protect\label{sec:Conclusion}Conclusion}
In summary, we have derived the orbital
magnetic moment of superconducting quasiparticles with a semiclassical approach. By constructing
a wavepacket with the Bogoliubov eigenstates, we calculate the energy of the wavepacket up to the linear order of the magnetic field and obtain the expression for the orbital magnetic moment. We verified the semiclassical result with a linear response approach. By comparing with the orbital angular momentum of the quasiparticle wavepacket, we reveal that the orbital magnetic moment has a more complicated structure. As a result, chiral superconductivity alone does not necessarily generate a nonzero orbital magnetic moment. We investigate the response of the superconducting quasiparticles due to the orbital magnetic moment, such as the magnetic modifications to the BdG band structure, corrections to the density of states, and the induced orbital Nernst effect. 
We apply our theoretical framework to a chiral 
$d$-wave superconductor on a honeycomb lattice. We show that the distribution of orbital magnetic moment can significantly deviate from the distribution of Berry curvatures, highlighting a
fundamental departure from electronic systems. We calculate the responses and present their temperature and chemical potential dependence.

In realistic superconductors, spin degrees of freedom may also contribute to transverse quasiparticle responses. For example, Ref.~\cite{Arxiv2024liao_semiSC} showed that the Berry curvatures of Bogoliubov quasiparticles can generate intrinsic spin Nernst and thermal Edelstein effects in spin-orbit-coupled superconductors. This suggests that a measured transverse response may contain both spin and orbital contributions. In the present work, we use a spin-degenerate model without spin-orbit coupling, so that the spin Nernst signal is absent and the orbital contribution can be isolated. For realistic materials with spin-orbit coupling or spin-triplet pairing, the spin response and the orbital magnetic moment response should be calculated within the same microscopic model and compared directly. Such a comparison can help determine whether an experimentally observed signal is mainly dominated by spin degrees of freedom or by the quasiparticle orbital magnetic moment.

We finally comment on possible material platforms. According to our formula, the quasiparticle orbital magnetic moment is sensitive to the structure of the normal-state Hamiltonian through the matrix element of $\tau_{z}\partial_{\mathbf{k}}\hat{H}_{\mathbf{k}}^{d}$. Promising platforms are therefore superconductors whose normal-state Hamiltonian contains suitable symmetry breaking or nontrivial internal structure. Rhombohedral multilayer graphene~\cite{Nat2025Han_RHG} and twisted bilayer $\rm{MoTe_2}$~\cite{Arxiv2025Xu_MoTe2} are natural examples, because their superconductivity appears close to normal states with broken time-reversal and inversion symmetries. Superconducting systems with Rashba spin-orbit coupling, such as Si(111)-($\sqrt{7}\times\sqrt{3}$)-In~\cite{NC2021Yoshizawa_Rashba} and Pb/Co/Si(111)~\cite{NC2017Gerbold_PbCoSi}, are also favorable because Rashba spin-orbit coupling introduces a momentum-dependent, off-diagonal internal structure into the normal-state Hamiltonian, which is a key ingredient for the orbital magnetic moment.

\section*{DATA AVAILABILITY}
The data that support the findings of this article are openly available~\footnote{\url{https://github.com/Jh-zeng-21/SC_OMM_Data}}.

\section*{Acknowledgments}
This work was supported by National Natural Science Foundation of China (Grant No. 12574153 and No. 12174453), the Guangdong Basic and Applied Basic Research Foundation (Grant No. 2026A1515010413), and the Guangdong Provincial Key Laboratory of Magnetoelectric Physics and Devices (Grant No. 2022B1212010008).

\appendix*
\section{Detailed derivation of the orbital magnetic moment within the semiclassical framework}

In this Appendix, we present a detailed derivation of the orbital magnetic moment of superconducting quasiparticles within the semiclassical wavepacket framework. We first construct the local BdG Hamiltonian in the presence of a slowly varying vector potential, and then evaluate the linear-in-field correction to the wavepacket energy. This correction defines the quasiparticle orbital magnetic moment. We finally obtain the expression in Eq.~(\ref{eq:OMMnumcal}), which is the central result used in the main text.

We start from the mean-field Hamiltonian of a superconducting system,
\begin{align}
\hat{H}
&=\int d\mathbf{r}\,c^{\dagger}(\mathbf{r})h_{0}(\mathbf{r},\mathbf{p}+e\mathbf{A}(\mathbf{r}))c(\mathbf{r}) \nonumber\\
&\quad+\iint d\mathbf{r}_{1}d\mathbf{r}_{2}
\left[\Delta(\mathbf{r}_{1},\mathbf{r}_{2})c^{\dagger}(\mathbf{r}_{1})c^{\dagger}(\mathbf{r}_{2})+\mathrm{h.c.}\right],
\label{eq:MFHamiltonianApp}
\end{align}
where $c^{\dagger}(\mathbf{r})$ creates an electron at position $\mathbf{r}$, $h_{0}$ is the normal-state electron Hamiltonian, $\mathbf{p}=-i\hbar\nabla_{\mathbf{r}}$ is the electron momentum operator, and $-e$ is the electron charge. The vector potential $\mathbf{A}(\mathbf{r})$ enters the electron Hamiltonian through the minimal coupling. The function $\Delta(\mathbf{r}_{1},\mathbf{r}_{2})$ denotes the superconducting gap between two electrons at $\mathbf{r}_{1}$ and $\mathbf{r}_{2}$.

External perturbations generally break the translational invariance of the crystal, which prevents a direct application of the band description. The semiclassical approach avoids this difficulty by describing the quasiparticle dynamics in terms of a wavepacket. This description is applicable when the external perturbations vary slowly on the scale of the quasiparticle wavepacket, which is typically set by the superconducting coherence length and is much larger than the Fermi wavelength. In this limit, the quasiparticle dynamics is governed by a local Hamiltonian $\hat{H}_{c}$, in which the external fields are evaluated at the wavepacket center $\mathbf{r}_{c}$. Since the spatial dependence of the external fields is locally frozen in this construction, $\hat{H}_{c}$ preserves the translational symmetry of the homogeneous system. The local BdG Hamiltonian is written as~\citep{Arxiv2024liao_semiSC}
\begin{equation}
\hat{H}_{c}=
\begin{pmatrix}
\hat{h}_{0}[\mathbf{r}_{c},\mathbf{r},\mathbf{p}+e\mathbf{A}(\mathbf{r}_{c})] & \hat{\Delta}(\mathbf{r}_{c},\mathbf{p})\\
\hat{\Delta}^{\dagger}(\mathbf{r}_{c},\mathbf{p}) & -\hat{h}_{0}^{*}[\mathbf{r}_{c},\mathbf{r},\mathbf{p}+e\mathbf{A}(\mathbf{r}_{c})]
\end{pmatrix}.
\label{eq:localBdGApp}
\end{equation}
Here the vector potential is chosen as the external perturbation used to study the orbital magnetic moment. The momentum dependence of the pairing Hamiltonian reflects the pairing symmetry associated with the relative position of the two electrons in a Cooper pair~\cite{Arxiv2024liao_semiSC}. Therefore, it does not depend on the vector potential.

The local Hamiltonian $\hat{H}_{c}$ satisfies the eigenvalue equation
\begin{equation}
\hat{H}_{c}|\psi_{n,\mathbf{k},\mathbf{r}_{c}}\rangle=E_{c}|\psi_{n,\mathbf{k},\mathbf{r}_{c}}\rangle .
\label{eq:localEigenApp}
\end{equation}
Because $\hat{H}_{c}$ is periodic in space, the quasiparticle eigenstate has the Bloch form $|\psi_{n,\mathbf{k},\mathbf{r}_{c}}\rangle=e^{i\mathbf{k}\cdot\mathbf{r}}|\phi_{n,\mathbf{k},\mathbf{r}_{c}}\rangle$, where $|\phi_{n,\mathbf{k},\mathbf{r}_{c}}\rangle$ is the cell-periodic part of the BdG eigenstate. In particle-hole space, it can be written as $|\phi_{n,\mathbf{k},\mathbf{r}_{c}}\rangle=(|u_{n,\mathbf{k},\mathbf{r}_{c}}\rangle,|v_{n,\mathbf{k},\mathbf{r}_{c}}\rangle)^{T}$, where $|u_{n,\mathbf{k},\mathbf{r}_{c}}\rangle$ and $|v_{n,\mathbf{k},\mathbf{r}_{c}}\rangle$ are the electron and hole components of the quasiparticle wave function, respectively.

A quasiparticle wavepacket in the $n$-th BdG band is constructed as~\citep{PRB1999Sundaram_semiclassical,PRB2017Liang_semiSC,Arxiv2024liao_semiSC}
\begin{equation}
|\Psi_{\mathbf{k}_{c},\mathbf{r}_{c}}\rangle=\int d\mathbf{k}\,\alpha_{\mathbf{k}}|\psi_{n,\mathbf{k},\mathbf{r}_{c}}\rangle ,
\label{eq:wavepacketApp}
\end{equation}
where $\alpha_{\mathbf{k}}=|\alpha_{\mathbf{k}}|e^{-i\gamma_{\mathbf{k}}}$ is the constructing function. The wavepacket is assumed to be sharply peaked around $\mathbf{k}_{c}$, with $\mathbf{k}_{c}=\int d\mathbf{k}\,|\alpha_{\mathbf{k}}|^{2}\mathbf{k}$.

The wavepacket energy is obtained by expanding the full Hamiltonian around the local Hamiltonian. To first order in the spatial gradient of the external field, one has
\begin{equation}
E\approx\langle\Psi|\hat{H}_{c}|\Psi\rangle+\langle\Psi|\Delta\hat{H}|\Psi\rangle=E_{c}+\Delta E_{n,\mathbf{k}_{c}},
\label{eq:wavepacketEnergyApp}
\end{equation}
where the gradient correction is~\citep{PRB1999Sundaram_semiclassical}
\begin{equation}
\Delta\hat{H}=\frac{1}{2}\left[(\hat{\mathbf{r}}-\mathbf{r}_{c})\cdot\frac{\partial\hat{H}_{c}}{\partial\mathbf{r}_{c}}+\frac{\partial\hat{H}_{c}}{\partial\mathbf{r}_{c}}\cdot(\hat{\mathbf{r}}-\mathbf{r}_{c})\right].
\label{eq:gradientCorrectionApp}
\end{equation}
Here $\hat{\mathbf{r}}=\mathbf{r}\tau_{0}$, and $\tau_{0}$ is the identity matrix in particle-hole space.

To extract the orbital magnetic moment, we consider a weak uniform magnetic field $\mathbf{B}$ and use the circular gauge $\mathbf{A}(\mathbf{r}_{c})=(\mathbf{B}\times\mathbf{r}_{c})/2$. We expand the local Hamiltonian $\hat{H}_{c}$ with respect to the vector potential $\mathbf{A}(\mathbf{r}_{c})$ and keep only the terms linear in $\mathbf{A}(\mathbf{r}_{c})$. This gives
\begin{equation}
\frac{\partial\hat{H}_{c}}{\partial\mathbf{r}_{c}}\approx \frac{e}{m}\frac{\partial[\mathbf{p}\cdot\mathbf{A}(\mathbf{r}_{c})]}{\partial\mathbf{r}_{c}}\tau_{0}=\frac{e}{2m}\mathbf{p}\times\mathbf{B}\,\tau_{0}.
\label{eq:HcGradientBApp}
\end{equation}
Substituting Eq.~(\ref{eq:HcGradientBApp}) into Eq.~(\ref{eq:gradientCorrectionApp}) gives
\begin{equation}
\Delta\hat{H}=\frac{e}{4m}\mathbf{B}\cdot\left[(\hat{\mathbf{r}}-\mathbf{r}_{c})\times\mathbf{p}\tau_{0}-\mathbf{p}\tau_{0}\times(\hat{\mathbf{r}}-\mathbf{r}_{c})\right].
\label{eq:DeltaHBApp}
\end{equation}
The first-order energy correction from a small homogeneous magnetic field can then be written as
\begin{equation}
\Delta E_{n,\mathbf{k}_{c}}=-\mathbf{B}\cdot\mathbf{m}_{n}(\mathbf{k}_{c}) ,
\end{equation}
which defines the orbital magnetic moment of the quasiparticle wavepacket as
\begin{equation}
\mathbf{m}_{n}(\mathbf{k}_{c})=-\frac{e}{4m}\langle\Psi|(\hat{\mathbf{r}}-\mathbf{r}_{c})\times\hat{\mathbf{p}}|\Psi\rangle+\mathrm{h.c.},
\label{eq:wavepacketMApp}
\end{equation}
where $\hat{\mathbf{p}}=\mathbf{p}\tau_{0}$ is the electron momentum operator acting on the particle-hole space.

We now evaluate Eq.~(\ref{eq:wavepacketMApp}). We first omit the Hermitian-conjugate part and calculate $\langle\Psi|\hat{\mathbf{r}}\times\hat{\mathbf{p}}|\Psi\rangle$. Using Eq.~(\ref{eq:wavepacketApp}), we find
\begin{align}
\langle\Psi|\hat{\mathbf{r}}\times\hat{\mathbf{p}}|\Psi\rangle
=\iint d\mathbf{k}' d\mathbf{k}\,\alpha^{*}_{\mathbf{k}'}\alpha_{\mathbf{k}}
\langle\phi_{n,\mathbf{k}'}|e^{-i\mathbf{k}'\cdot\mathbf{r}}\hat{\mathbf{r}}\times\hat{\mathbf{p}}e^{i\mathbf{k}\cdot\mathbf{r}}|\phi_{n,\mathbf{k}}\rangle .
\label{eq:packetAverageStartApp}
\end{align}
Introducing $\hat{\mathbf{p}}_{\mathbf{k}}=e^{-i\mathbf{k}\cdot\mathbf{r}}\hat{\mathbf{p}}e^{i\mathbf{k}\cdot\mathbf{r}}$ and using $\hat{\mathbf{r}}e^{-i(\mathbf{k}'-\mathbf{k})\cdot\mathbf{r}}=i\partial_{\mathbf{k}'}e^{-i(\mathbf{k}'-\mathbf{k})\cdot\mathbf{r}}$, one can integrate by parts over $\mathbf{k}'$~\citep{PRB1996Chang_electronOMM}. With $\alpha_{\mathbf{k}}=|\alpha_{\mathbf{k}}|e^{-i\gamma_{\mathbf{k}}}$ and the narrow wavepacket approximation, the result becomes
\begin{align}
\langle\Psi|\hat{\mathbf{r}}\times\hat{\mathbf{p}}|\Psi\rangle
=&\bigg[\frac{i}{2}\partial_{\mathbf{k}}\times\langle\phi_{n,\mathbf{k}}|\hat{\mathbf{p}}_{\mathbf{k}}|\phi_{n,\mathbf{k}}\rangle 
+\partial_{\mathbf{k}}\gamma_{\mathbf{k}}\times\langle\phi_{n,\mathbf{k}}|\hat{\mathbf{p}}_{\mathbf{k}}|\phi_{n,\mathbf{k}}\rangle \nonumber\\
&-i\langle\partial_{\mathbf{k}}\phi_{n,\mathbf{k}}|\times\hat{\mathbf{p}}_{\mathbf{k}}|\phi_{n,\mathbf{k}}\rangle\bigg]_{\mathbf{k}=\mathbf{k}_{c}} .
\label{eq:packetAverageFinalApp}
\end{align}
Here and in the following, matrix elements involving the cell-periodic states $|\phi_{n,\mathbf{k}}\rangle$ are understood as inner products over the unit cell.

The center-position contribution in Eq.~(\ref{eq:wavepacketMApp}) is
\begin{equation}
\mathbf{r}_{c}\times\langle\Psi|\hat{\mathbf{p}}|\Psi\rangle
=\left[\mathbf{r}_{c}\times\langle\phi_{n,\mathbf{k}}|\hat{\mathbf{p}}_{\mathbf{k}}|\phi_{n,\mathbf{k}}\rangle\right]_{\mathbf{k}=\mathbf{k}_{c}} .
\label{eq:centerContributionApp}
\end{equation}
The wavepacket center satisfies $\mathbf{r}_{c}=\partial_{\mathbf{k}_{c}}\gamma_{\mathbf{k}_{c}}+\mathbf{A}_{\mathbf{k}_{c}}$~\citep{PRB1999Sundaram_semiclassical,PRL2021Wang_semiSC,Arxiv2024liao_semiSC}, where $\mathbf{A}_{\mathbf{k}}=i\langle\phi_{n,\mathbf{k}}|\partial_{\mathbf{k}}\phi_{n,\mathbf{k}}\rangle$ is the quasiparticle Berry connection. Combining Eqs.~(\ref{eq:packetAverageFinalApp}) and (\ref{eq:centerContributionApp}), and then restoring the Hermitian-conjugate term, we obtain
\begin{align}
\mathbf{m}_{n}(\mathbf{k}_{c})
&=-\frac{e}{4m}\bigg[
\left(\frac{i}{2}\partial_{\mathbf{k}}-\mathbf{A}_{\mathbf{k}}\right)
\times\langle\phi_{n,\mathbf{k}}|\hat{\mathbf{p}}_{\mathbf{k}}|\phi_{n,\mathbf{k}}\rangle \nonumber\\
&\quad -i\langle\partial_{\mathbf{k}}\phi_{n,\mathbf{k}}|\times\hat{\mathbf{p}}_{\mathbf{k}}|\phi_{n,\mathbf{k}}\rangle+\mathrm{h.c.}\bigg]_{\mathbf{k}=\mathbf{k}_{c}} .
\label{eq:OMMBeforeCovariantApp}
\end{align}
This expression can be simplified into
\begin{equation}
\mathbf{m}_{n}(\mathbf{k}_{c})=-\frac{e}{2m}\mathrm{Re}\left[
\langle D_{\mathbf{k}}\phi_{n,\mathbf{k}}|\times\hat{\mathbf{p}}_{\mathbf{k}}|\phi_{n,\mathbf{k}}\rangle
\right]_{\mathbf{k}=\mathbf{k}_{c}},
\label{eq:OMMCovariantApp}
\end{equation}
where $D_{\mathbf{k}}\equiv i\partial_{\mathbf{k}}-\mathbf{A}(\mathbf{k})$ is the gauge-invariant $\mathbf{k}$-space derivative. From this formula we see that the orbital magnetic moment of each wavepacket depends only on its momentum-space center ${\mathbf k}_c$. In the following, we omit the subscript $c$ of $\mathbf{k}_{c}$ for simplicity.

The momentum operator entering the magnetic coupling is related to the diagonal part of the BdG Hamiltonian by
\begin{equation}
\hat{\mathbf{p}}_{\mathbf{k}}=\frac{m}{\hbar}\tau_{z}\partial_{\mathbf{k}}\hat{H}^{d}_{\mathbf{k}},
\label{eq:pkHdRelationApp}
\end{equation}
where $\hat{H}^d_{{\mathbf k}} ={\rm diag}(e^{-i\mathbf{k}\cdot\mathbf{r}}\hat h_0 e^{i\mathbf{k}\cdot\mathbf{r}}, -e^{-i\mathbf{k}\cdot\mathbf{r}}\hat h^*_0 e^{i\mathbf{k}\cdot\mathbf{r}})$. The Pauli matrix $\tau_{z}$ appears because the electron and hole components carry opposite charges.

Substituting Eq.~(\ref{eq:pkHdRelationApp}) into Eq.~(\ref{eq:OMMCovariantApp}) and inserting the completeness relation $\sum_{n'}|\phi_{n',\mathbf{k}}\rangle\langle\phi_{n',\mathbf{k}}|=1$, one can separate the intraband and interband contributions. The intraband part is canceled by the Berry connection term in Eq.~(\ref{eq:OMMCovariantApp}), leaving only terms with $n'\neq n$. For these interband terms, differentiating the eigenvalue equation $\hat{H}_{\mathbf{k}}|\phi_{n,\mathbf{k}}\rangle=E_{n,\mathbf{k}}|\phi_{n,\mathbf{k}}\rangle$ gives
\begin{equation}
\langle\partial_{\mathbf{k}}\phi_{n,\mathbf{k}}|\phi_{n',\mathbf{k}}\rangle
=
\frac{\langle\phi_{n,\mathbf{k}}|\partial_{\mathbf{k}}\hat{H}_{\mathbf{k}}|\phi_{n',\mathbf{k}}\rangle}{E_{n,\mathbf{k}}-E_{n',\mathbf{k}}},
\qquad n'\neq n .
\label{eq:interbandIdentityApp}
\end{equation}
Using Eq.~(\ref{eq:interbandIdentityApp}), the orbital magnetic moment is finally obtained as
\begin{align}
\mathbf{m}_{n}(\mathbf{k})&=\frac{e}{2\hbar}\sum_{n^{\prime}\neq n}\mathrm{Im}\left[\frac{\langle\phi_{n}|\partial_{\mathbf{k}}\hat{H}_{\mathbf{k}}|\phi_{n^{\prime}}\rangle\times\langle\phi_{n^{\prime}}|\tau_{z}\partial_{\mathbf{k}}\hat{H}^d_{\mathbf{k}}|\phi_{n}\rangle}{E_{n',\mathbf{k}}-E_{n,\mathbf{k}}}\right].
\label{eq:OMMnumcalApp}
\end{align}
Here $\hat{H}_{\mathbf{k}}=e^{-i\mathbf{k}\cdot\mathbf{r}}\hat{H}_{c}e^{i\mathbf{k}\cdot\mathbf{r}}$ is the $\mathbf{k}$-dependent BdG Hamiltonian. Eq.~(\ref{eq:OMMnumcalApp}) is the central result for the orbital magnetic moment of superconducting quasiparticles.

\bibliography{omm}

\end{document}